\documentclass[preprint,aps,eqsecnum,axodraw]{article}
\usepackage{feynmf}
\usepackage{graphicx}
\usepackage{amsbsy}

\newcommand{\beq}{\begin{eqnarray}}
\newcommand{\eeq}{\end{eqnarray}}

\def\lsim{\mathrel{\rlap{\lower4pt\hbox{\hskip1pt$\sim$}}
    \raise1pt\hbox{$<$}}}         
\def\gsim{\mathrel{\rlap{\lower4pt\hbox{\hskip1pt$\sim$}}
    \raise1pt\hbox{$>$}}}         

\def\overseta#1#2{\rlap{ \hspace{.3cm}
\raise 13pt \hbox{$#1$}}{\lower 2pt\hbox{{\huge $#2$}}}}

\def\oversetb#1#2
{\rlap{\hspace{-0.1cm}\raise 6pt \hbox{$#1$}}
{\lower 3pt\hbox{{\huge $#2$}}} }

\def\underset#1#2{\ \rlap{\lower 3pt \hbox{$#1$}}{\raise 2pt \hbox{$#2$}}\ }

\begin{document}
 
\begin{fmffile}{paper14pics}
\begin{titlepage}
\vskip -.6in
\flushright  \vspace*{-1.5cm}{\small BNL-HET-04/13, LBNL-55325}
\vskip .35in

\begin{center}
{\large \bf Flavor Structure of Warped Extra Dimension Models}
\vskip .15in
Kaustubh Agashe \footnote{email: kagashe@pha.jhu.edu} 
\vskip .1in
{\em Department of Physics and Astronomy \\
Johns Hopkins University \\
Baltimore, MD 21218-2686} \\
\vskip .2in
Gilad Perez \footnote{email: Gperez@lbl.gov} 
\vskip .1in
{\em Theoretical Physics Group \\ Lawrence Berkeley National Laboratory \\ Berkeley, CA 94720}
\vskip .2in
and \\
\vskip .2in
Amarjit Soni \footnote{email: soni@bnl.gov}
\vskip .1in
{\em High Energy Theory Group \\ Brookhaven National Laboratory \\ 
Upton, New York 11973}

\end{center}
\vskip .05in

\begin{abstract}

We recently showed, in hep-ph/0406101, that warped
extra dimensional models with bulk custodial symmetry and
few TeV KK masses 
lead to striking signals at $B$-factories.
In this paper, using a spurion analysis, we 
systematically study the 
flavor structure of models 
that
belong to the above class. 
In particular we find that the profiles of the zero modes,
which are similar in all these models,
essentially control the underlying flavor structure.
This implies that our results are robust
and model independent in this class of models.
We 
discuss in detail the origin of the signals 
in B-physics.
We also briefly study other NP signatures
that arise in rare K decays ($K \to \pi \nu \nu$),
in rare top decays [$t \to c \gamma (Z, gluon)$] and  
the possibilty of CP
asymmetries in $D^0$ decays to CP eigenstates
such as $K_S \pi^0$ and others.   
Finally we 
demonstrate that with light KK masses, $\sim3$ TeV, the above
class of models with anarchic $5D$ Yukawas has a
``CP problem'' since contributions to the neutron electric dipole
moment  
are  
roughly 20 times larger
than the current experimental bound. 
%
%
Using AdS/CFT
correspondence, these extra-dimensional models are dual to a
purely $4D$ strongly
coupled conformal Higgs sector thus enhancing their appeal.

\end{abstract}

\end{titlepage}

\newpage
\renewcommand{\thepage}{\arabic{page}}
\setcounter{page}{1}
\section{Introduction}
The Standard Model (SM) has been very successful so far. Almost
all of its predictions that have been tested were verified to high
precision. Nevertheless, the SM raises the fine tuning problem -
which is related to the smallness of the ratio $\Lambda_{\rm
  EWSB}/M_{\rm Pl}$, where $\Lambda_{\rm EWSB}$
is the electroweak symmetry breaking  (EWSB) scale and $M_{\rm Pl}$ is
the Planck mass. The
Randall-Sundrum model (RS1), 
with a warped extra dimension (WED), provides a natural solution to the
above problem~\cite{rs1}:
due to warping, there is 
exponential hierarchy between the effective cut-off scales of the
theory at the two ends of the extra dimension. 
Thus $\Lambda_{\rm EWSB}$ is protected
due to a low cut-off near the TeV brane while the high
scale of gravity is generated at the other end.

In the original RS1 model the SM fields live on the TeV brane.
Consequently, the model cannot solve the
flavor puzzle, {\it i.e.}, why are most of the flavor parameters
small and hierarchical. Furthermore, higher dimension operators
that induce contribution to FCNC processes are suppressed only by powers
of $\Lambda_{\rm EWSB}\,$ (the effective cut-off).
As well known, however, the cut-off required
to suppress $\epsilon_K$ and other observables should be higher than
${\cal O}(1000)\,$TeV. 
The coefficients of these higher-dimensional operators, which cannot
be calculated from first principles,
can be in fact small. Consequently, in the original RS1 framework
flavor issues are UV
sensitive.
Similar arguments apply also to constraints from electroweak precision
measurements (EWPM) which required cut-off higher than ${\cal O}(10)\,$TeV. 

In the sense of AdS/CFT correspondence \cite{adscft}, 
this RS1 model is dual \cite{rscft} to a $4D$ CFT of which
the Higgs is a composite which arises
after conformal invariance is broken, {\it i.e.}, RS1 is dual to  
a strongly interacting Higgs sector model.
Thus, the hierarchy problem is solved by compositeness of Higgs.
In addition, the SM gauge and fermions fields are also composites so that 
flavor issues 
(and corrections to electroweak precision observables) depend 
on details of this compositeness (dual to UV sensitivity
on RS1 side). In general, we expect FCNC's suppressed
by compositeness scale of $\sim$ TeV with{\em out} any small
coefficients and thus too large.

An alternative to the above is that only Higgs is a composite of a strongly
interacting sector \cite{comphiggs}, in this case a CFT, while 
the SM gauge and fermions fields are fundamental fields, external to
the CFT. This suffices to solve the hierarchy problem since only the Higgs
mass requires protection, the masses of gauge and fermion fields being protected by 
symmetries.
However, for gauge boson and fermion masses to arise at 
the weak scale, these fields must couple to the CFT/Higgs sector.
The RS dual of this $4D$ set-up is SM gauge and fermion fields in the bulk.

The above UV sensitivity is removed when the fermions and the gauge
bosons are allowed to propagate in the bulk.
One then uses the idea of split fermions by localizing the
light quarks near the Planck brane.
This modification of the model yields three important virtues:
\begin{itemize}
\item[(i)] Automatic suppression of the higher dimensional operators 
due to a high effective cut-off near Planck brane \cite{gp, huber}. 
\item[(ii)] The
flavor puzzle is ameliorated, since the quark masses and mixing 
are determined by
the value of their wave function (WF) on the TeV 
brane~\cite{neubert, gp, huber}.
\item[(iii)] Since the SM fields are in the bulk presence of new KK
states is implied.
The coupling of these new states are both custodial and flavor symmetry
violating. It turns out however (as we explain below) that 
flavor dependence in 
the
couplings
between these new heavy states and the light fermion modes are suppressed
(unlike as in the flat extra dimension case) by
combination  of
RS-GIM mechanism and approximate symmetries.
\end{itemize}

Constraints from EWPM in the above setup were shown to
imply a bound on the KK masses $m_{\rm KK}\gsim{\cal O}(10)\,$TeV~\cite{EWPM}.
This bound, 
combined with
virtue (iii), 
makes the model 
consistent with
constraints from FCNC processes~\cite{huber}.
In spite of this success the model now raises the little hierarchy problem:
Fine-tuning is required to explain the smallness of the
EWSB scale,
$\Lambda_{\rm EWSB}$, 
compared to the lightest non-zero KK mass,
$$\left({\Lambda_{\rm EWSB}\over m_{\rm KK}}
\right)^2 
={\cal O}\left(10^{-4}\right)\,.$$

Reference~\cite{cust1} (for models with a Higgs) and~\cite{cust2}
(for Higgsless models)
improved upon the above
by promoting the gauge symmetry in the bulk to be custodial
invariant.
It was shown in~\cite{cust1} that 
a model with KK masses of ${\cal O}(3\,{\rm TeV})$ can be consistent
with EWPM for the Higgs case.
The consistency with EWPM for the Higgsless models was considered
in~\cite{barbieri1, hewett1,nomura2,ccgt,barbieri2,hewett2} 
with partial success.
In any case, one may expect that 
the little hierarchy problem would be reduced to the 
O(1\%) bearable range 
in model with Higgs on TeV brane and to the O(10\%)
range in models
with Higgs in the bulk (but with a wavefunction
peaked near TeV brane) \cite{bulkhiggs};
whereas, there is no fine-tuning in Higgsless models.
%

\subsection{Overview}
As discussed above the  way the zero mode profiles are located in
the extra dimension plays an essential role in the success of the above models.
This implies that up to some limited freedom the flavor parameters
of the framework are fixed.
Thus our aim below is to make a systematic study of the structure of flavor violation in
this framework. A discussion of the signals for B-physics already
appeared in our previous work \cite{Agashe:2004ay}.

We perform our analysis under the {\it assumption} that
the model contains the minimal amount of fine tuning and hierarchies
in its fundamental parameters.
In particular we {\it assume} low KK masses, $$m_{\rm KK}\lsim 3\,{\rm TeV}\,,$$ and
also that the entries in the 5D Yukawa matrices are complex and of the
same order.
Since the KK masses are smaller than in earlier studies,
we expect the FCNC's to be enhanced leading to non-trivial constraints and signals.
%
%
An earlier study of flavor violation with few TeV KK masses appears in
reference \cite{Burdman:2003nt}. However, as we will discuss later, 
hierarchies in $5D$ Yukawa were 
allowed in that study leading to quite different
conclusions/signals than in our study.

Having made these assumptions, we 
then look for signals which will be able
to test the model's predictions related to the flavor sector.
As it turns out, the structure of flavor violation in 
the KK theory 
mostly depends on value of zero mode profiles near the TeV brane. 
These are expected to be similar
in all the existing models. Consequently our results are robust and 
independent of details within this class of models.

As mentioned above, there is a combination of RS-GIM mechanism 
and approximate symmetries
for light fermions leading to suppressed FCNC's.
We show below that, just as in the SM, the 
RS-GIM mechanism is violated by large top quark mass.
%
This result in the following three types 
of new physics (NP) contributions:  
%
\begin{itemize}
\item[(I)] Contributions to $\Delta F=2$ FCNC processes - arise from a
  relative large dispersion in the doublets 5D-masses, 
specifically large
coupling of $(t,b)_L$ to gauge KK modes due to heaviness of the top quark
  \item[(II)] Contributions to $\Delta F=1$ FCNC processes (mostly semileptonic) - 
arise
    from  combination of (I) and mixing between the zero and KK states of
    the $Z$ due to EWSB.
    \item[(III)] Contributions to radiative B decay (dipole operators)
      - arise from 
large $5D$ Yukawa required to obtain top quark mass combined with 
large mixing in the right handed down type
      diagonalization matrix, $D_R\,.$
  \end{itemize}
  In addition we also discuss the NP contributions to electric dipole
  moments (EDM)s. We find that 1-loop contributions are
predicted to be of $O(10)$ larger than
the current experimental sensitivity for 
$m_{ KK } \stackrel{<}{\sim}4 $ TeV which implies an RS CP problem. 

In section~\ref{RSCus} we briefly introduce the framework and
describe the models we consider.
Section~\ref{FlSt} discusses the flavor structure of the framework and
presents a spurion analysis for the KK theory. This makes 
the structure of flavor violation in the theory more
transparent and easy to handle. In section \ref{cft},
we give an interpretation of the flavor structure in
the dual CFT.
In section~\ref{sig} we consider the predictions of our framework for
various
observables related to the above underlying flavor structure.

In particular, in
subsection~\ref{DF2} we show that the models predict
sizable $\Delta F=2$ NP contributions 
which leads to what we denote as the flavor ``coincidence'' problem.
In subsection~\ref{DF1} we discuss $\Delta F=1$ contributions and argue
the the NP contribution are likely to be observed only 
in semi-leptonic decays (e. g. $B \to X_s l^+ l^-$).
In subsection~\ref{Dipole} we estimate the NP contributions to various
radiative B decays.
In subsection~\ref{EDM} we consider the NP contributions to 
EDMs which are related to
flavor diagonal CPV. In subsection~\ref{up} we briefly comment about
flavor
violation in the up type sector.
Finally we conclude in section~\ref{Conc}.
Appendices contain technical details.

\section{The framework}
\label{RSCus}
We begin with a  description of the model independent features of
the framework under study~\cite{cust1,cust2,nomura1,csaki}.
We shall then briefly comment about the differences between
the relevant models considered below.
The basic set-up of our models is the RS1 framework~\cite{rs1}.
The space time of the model is described by 
a slice of ADS$_5$ with curvature scale, $k\sim M_{ Pl }$, the $4D$ Planck mass.
The 
Planck brane is located at $\theta=0$, where $\theta$ is the compact extra dimension coordinate.
The TeV brane is located at 
$\theta = \pi$. 
The metric of RS1 can be written as:
\begin{equation}
( d s) ^2  = \frac{1}{ ( kz )^2 } \big[ \eta_{ \mu \nu }
d x ^{ \mu } d x^{ \nu } - ( d z ) ^2 \big]\,, 
\end{equation}
where $kz=e^{kr_c\theta}\,.$
  We assume that
$k \pi r_c \sim \log \left( M_{\rm Pl} / {\rm TeV} 
\right)$
to solve the hierarchy problem,
\begin{equation}
\left( z_h \equiv \frac{1}{k} \right) 
\leq  z \leq \left( z_v \equiv \frac{ e^{ k \pi r_c }}{k} \right),
\end{equation}
where $z_v \sim \mbox{TeV}^{-1}$. 

The gauge group of the models under study is given by~\cite{cust1,cust2}
$SU(3)_c \times SU(2)_L \times SU(2)_R \times
U(1)_{ B - L }$.
The gauge symmetry is broken on the Planck brane down to the SM gauge group and
in the TeV brane it is broken down to $SU(3)_c \times SU(2)_D \times U(1)_{ B - L }$. 
$SU(2)_D$ is the diagonal subgroup of the two $SU(2)$'s 
present in the bulk.

The fermion and the scalar fields content is model dependent.
We shall elaborate more on the fermions sector in the following
section.
The major role played by the Higgs field, 
relevant to our consideration, is to yield 
the masses and mixing for the SM fields and in addition
mixing between these and the higher KK fields.
In that sense it is not important whether we consider the
Higgs~\cite{cust1}
or Higgsless models~\cite{cust2,nomura1,csaki}.
Thus we will not elaborate more on this subject. In points, however,
in which the difference between the models is relevant we shall
explicitly specify that. 

The Lagrangian of the models can be described by
\begin{equation}
S=\int d^4x dz \sqrt{G}\left[
{\cal{L}}_{gauge}+
{\cal L}_{fermion}+ 
{\cal L}_{\rm{Pl}}\delta\left(z-z_h\right)+{\cal L}_{\rm{TeV}}
\delta\left(z-z_v\right)\right]\,,
\end{equation}
where
${\cal L}_{gauge} + {\cal L}_{fermion}$ is the bulk Lagrangian.
The bulk gauge Lagrangian is
\begin{eqnarray}
{\cal L}_{ gauge } = \sqrt{G} \left( 
- \frac{1}{ 4 \left(g_{ 5D }^i\right)^2 } F_{ i \; MN } F_i^{ MN } \right)
\end{eqnarray}
where $g_{ 5D }$ is the $5D$ gauge coupling
and the index $i$ runs over the $SU(3)_c$, $SU(2)_L$, $SU(2)_R$ and
$U(1)_{ B - L }$ gauge fields.
The fermion Lagrangian will be presented in the next section.

The TeV brane Lagrangian contains the EWSB sector (including SM fermion
mass terms: see Eq. (\ref{LfermionTeV})).
The Planck brane Lagrangian, ${\cal L}_{ Pl }$ contains 
necessary fields to spontaneously break bulk gauge symmetry
to $SU(2)_L \times U(1)_Y$ and also to break degeneracy between
up and down quark masses in some models (see below).


\subsection{Fermions}

The fermion sector of \cite{cust1,nomura1} is described as:
$Q=(3,2,1)_{1\over3},u=(3,1,2)_{1\over3},d=(3,1,2)_{1\over3}\,,$
where the number in the parenthesis stands for the fermion
representation
under the $SU(3)_c \times SU(2)_L \times SU(2)_R \times
U(1)_{ B - L }$ gauge group respectively, and all of them propagate in
the bulk.
We use the notation $Q,u,d$ to match with the transformation of the
light modes, $Q^0,u^0,d^0$, belonging to the above fields under the SM
gauge group.~\footnote{We use the terms zero modes and light modes for
the SM fields interchangeably.}
Thus $Q$ contains two SM-light fields and $u,d$ contain each a single
SM-light field.
The fermion sector of \cite{csaki} consist of 
$Q=(3,2,1)_{1\over3},\,Q_R=(3,1,2)_{1\over3}\,,$ where in that case
$Q_R$ contains both the SM up and down type singlets.
In addition, to break the degeneracy between the two light components
of $Q_R$ a pair of Planck brane vector like quarks was added.
These couple to the up type singlet component of $Q_R$.
Consequently its WF is distorted and splitting between the up and down SM singlet quarks is
achieved.

At the end of the day, one finds that to account for the SM masses and mixing the
profile of the zero modes in~\cite{csaki} is similar to the ones
in~\cite{cust1,nomura1}. This is one of the main reasons for the fact
that our analysis is robust and model independent.
In order to do actual calculation we arbitrarily
choose to work
with the fermion sector
of~\cite{cust1,nomura1}. We nevertheless have in mind that our analysis can be
straightforwardly translated into the language of~\cite{csaki}.


\section{Flavor Structure}
\label{FlSt}

\subsection{Flavor violation - 5D theory}\label{FV5D}

%
The relevant piece of the Lagrangian, related to the flavor sector, is given by
a bulk piece
\begin{equation}
{\cal L}_{fermion} = \sqrt{G} \left[ 
i \bar{ \Psi _i } \Gamma^M D_M \Psi_i + k C_{Q,u,d}
 \left(\bar{Q},\bar u,\bar d\right)\left({Q},u,d \right)
\right]
\end{equation}
(the index $i$ runs over the different fermion
representations)
and a TeV-brane localized piece:
\begin{equation}
{\cal L}_{ TeV } \ni
h\delta\left(z-z_v\right)\lambda_{u,d\, 5D}\bar Q\left(u,d\right) \,,\label{L5DF}
\label{LfermionTeV}
\end{equation}
where for the Higgs models $h$ stands for the Higgs field and for the
Higgsless ones it stands for a mass parameter $h\to\langle
h\rangle=v\simeq 174\,$GeV. In addition 
$C_{Q,u,d}$ are $3\times 3$ hermitian matrices (due to 5D Lorentz
symmetry) and $\lambda_{u,d\, 5D}$ are the 5D Yukawa matrices.
In principle, since the above theory is non-renormalizable, higher
dimension, flavor violating, operators should be present
in~(\ref{L5DF}). Due to the fact that the light quarks are
localized near the Planck brane the effective cut-off relevant to this
type of operators is very high. Thus they are subdominant and are
neglected in our analysis below.

Unlike the SM case, in addition to the Yukawa matrices
the model contains also additional sources of flavor violation in
the form of $C_{Q,u,d}\,.$
 $${\rm U(3)}_{Q,u,d}\, \ 
    {\overseta{{{\hspace*{-.54cm}C_{Q,u,d}}}}{\longrightarrow}} \, \ {{\rm U(1)}^3}_{Q,u,d}\,,$$
  where U(3)$_{Q,u,d}$ is the flavor group of the 5D
  theory, per representation, in the limit where
  $C_{Q,u,d},\lambda_{u,d\, 5D}\to0\,.$
  Indeed as discussed below the addition of these matrices induces
  the non-trivial flavor structure of this framework.

  One can count the number of the physical flavor parameters
  present in the above theory. Generically, however, there is no
  direct translation between the flavor parameters
  of the 5D theory and the ones appearing in the IR limit of the theory.
  This is the case since it depends on integrals of 
$z$
dependent
  functions.
  Thus the quantities which characterize the flavor violation in the
  4D effective theory are functionals of $C_{Q,u,d}\,.$~\footnote{This
  is generically the situation in any 5D theory which realizes the
  split fermions idea.}
  In the RS framework, as discussed below, there is, in fact, a simple relation between the
  structure of flavor violation in the 5D and 4D theories.

  Thus it is worthwhile to do the above counting.
  For reasons that will be clear below it is useful to separate
  the counting into the following two cases.
  We start with the case in which flavor violations occur only in
  the down type quark sector, $C_{u},\lambda_{u\, 5D}\to0\,.$
  Then we consider the generic case.~\footnote{For more details on
  counting flavor parameters see {\it e.g.} \cite{Nir} and refs therein.}
  \begin{itemize}
  \item[(i)] Down type quark sector - $\lambda_{d\, 5D}$ contains 9
    real and 9 imaginary parameters and $C_{Q,d}$ contains, each, 6
    real and 3 imaginary parameters. Altogether we find 21 real and
    15 imaginary parameters.
    The $U(3)_{Q,d}$ global symmetries can be use to
    eliminate, however, 6 real and 11 phases which are unphysical
  (one phase cannot be removed since
    it corresponds to an unbroken baryon number symmetry.).
    Thus after applying the above rotation of the fields we find
    15 real and 4 imaginary physical flavor parameters. This case is
    summarized
    in table~\ref{down}.
  \item[(ii)] Generic model - $\lambda_{u,d\, 5D}$ contains 18
    real and 18 imaginary parameters and $C_{Q,u,d}$ contains 18
    real and 9 imaginary parameters. Altogether we find 36 real and
    27 imaginary parameters. The $U(3)_{Q,u,d}$ 
    symmetries can be used to
    eliminate 9 real and 17 phases.
    Thus, we find
    27 real and 10 imaginary physical flavor parameters. This case is
    summarized
    in table~\ref{Tot}.
  \end{itemize}
  The physical role of these parameters is obscure at this level.
  We shall see, however, that in the KK theory the role of the above
  parameters is more transparent.

\begin{center}
   \begin{table}[!hb]\begin{center}
\begin{tabular}{||c|c|c||}
\hline\hline
&Re& Im\\ \hline  
$\lambda_{d\, 5D}$&9&9\\
\hline
$c_{Q}+c_d$&12&6\\
\hline
Total&21&15\\
\hline
Non-phys' parameters& 6&11\\ \hline
Phys' parameters&15 &4 \\
\hline\hline
\end{tabular}
\caption{\small Number of flavor parameters for a theory with only down type
  quark sector.}
\label{down}\end{center}                               
 \end{table}\end{center}


   \begin{table}[!ht]\begin{center}
  \begin{tabular}{||c|c|c||}
\hline\hline
&Re& Im\\ \hline  
$\lambda_{u\, 5D}+\lambda_{d\, 5D}$&2$\times$9&2$\times$9\\
\hline
$c_{Q}+c_u+c_d$&3$\times$6&3$\times$3\\
\hline
Total&36&27\\
\hline
Non-phys' parameters& 9&17\\ \hline
Total phys' parameters&27 &10 \\
\hline\hline
\end{tabular}
\caption{\small Total number of flavor parameters in the full theory.}
\label{Tot} \end{center}                              
 \end{table}

\subsection{Flavor violation - KK theory}
\subsubsection{Zero modes and SM flavor parameters}
The fermion zero-modes are identified with the observed SM fermions.
As explained in the introduction our working assumption is that the 
5D Yukawa matrices are anarchical.
The hierarchy in
SM flavor parameters is, therefore, directly related to the zero mode
profiles in the 5D through the split fermion mechanism.
In order to see how this works we start with considering
only the zero modes sector of our framework.

We can always do actual computations in a basis in which the bulk
masses are diagonal, $C_{Q,u,d}={\rm diag}\left(c_{Q^i,u^i,d^i}\right)\,.$
In this basis, we shall denote as the ``special basis''
the WF of the, canonically normalized, zero-mode fermions are given 
by (see, for example, \cite{huber})
\begin{equation}
 \hat\chi_0 (c_{x^i}, z)\equiv z^{-{3\over2}}\chi_0 (c_{x^i}, z) = \sqrt{ 
\frac{k\left(1-2c_{x^i}\right)}{ \left( kz_v\right)^{ 1-2c_{x^i} }
-1  } }\, \left( kz \right)^{{1\over2}-c_{x^i}}\,,
\label{zero}
\end{equation}
where $\chi_0 (c_{x^i}, z)$ is the zero mode profile, 
$x=Q,u,d\,,$ and in the above
we neglected electroweak breaking effects on the TeV brane.
It is evident from (\ref{zero}) that when
$c_{x^i} > 1/2$ $(c_{x^i} < 1/2)$ the zero-mode 
fermion is localized
near Planck (TeV) brane. 

Using (\ref{L5DF}) we find that the effective 4D Yukawa matrices $\lambda_{u,d \; 4D }$
are given by:
\begin{equation}
\lambda_{u,d\,4D}^{ij} = 
\frac{ 2 \lambda_{u,d\, 5D}^{ij} k }{ f_{ Q^i}  
f_{u^j,d^j} }\,,
\label{lambda4D}
\end{equation}
where
\begin{equation}
{2k\over f_{x^i}^2}=  \hat\chi_0^2\left(c_{x^i}, z_v\right)
\,.
\label{fdef}
\end{equation}
It is useful to find the asymptotic dependence of $f_{x^i}$
on $c_{x^i}$
\begin{eqnarray}
f_{x^i}^{-2} \sim \left\{
\begin{array}{ll}
 \frac{1}{2} - c_{x^i} & \,\mbox{for}\,\ c_{x^i} < {1\over2} - \epsilon  \\  \\
 \frac{1}{ 2 k \pi r_c }&\, \mbox{for}\,\ c_{x^i}\to{1\over2}\\ \\
  \left(c_{x^i}-\frac{1}{2}\right)e^{k \pi r_c \left( 
1 -2 c_{x^i} \right) } & \,\mbox{for}\,\
 c_{x^i} > {1\over2} + \epsilon, 
\label{fapp}
\end{array}
\right.
\end{eqnarray}
where $\epsilon \sim 0.1$ so that the asymptotic value of $f_{x^i}$
is obtained rapidly for $c > (<) 1/2$. 

Generically $ \lambda_{u,d\, 5D}$ are expected to be anarchical. We
therefore assume that all the entries in these Yukawa matrices are
complex and of order unity.
As a consequence the hierarchy in the SM flavor parameters
should be accounted for by the corresponding values of $f_{Q,u,d}\,.$
Up to an overall dimensionful proportionality coefficient, $\lambda_{5D}$,
 the following relation between $f_{Q,u,d}\,,$ and the flavor
 parameters should hold,
 \beq
m_{u^i,d^i}\sim {2 v\lambda_{5D}k \over f_{Q^i} f_{u^i,d^i}}
\,,\label{4Dm}
\eeq
where $v\simeq174\,$GeV.
Furthermore, the size of the elements of $U_{L,R}$, ($D_{R,L}$)
the diagonalization matrices of the up (down) Yukawa matrices (related
to the SM doublet and singlets field respectively) are given by:
\beq
&&\left|\left(D_L\right)_{ij}\right|\sim\left|\left(U_L\right)_{ij}\right|\sim
\left|\left(V_{\rm CKM}\right)_{ij}\right|\sim {f_{Q^i}\over
  f_{Q^j}}\, \ \ \ {\rm for} \ \ \ j\leq i\,,
\nonumber\\ 
&&\left|\left(U_R,D_R\right)_{ij}\right|\sim {f_{u^i,d^i}\over
  f_{u^j,d^j}}\, \ \ \ {\rm for} \ \ \ j\leq i\,,
\label{fs}
\eeq
where $ V_{\rm CKM}\equiv U_L^\dagger D_L\,,$ and in the above for
$j>i$ one should interchange the $i$ and $j$ indices.


\subsubsection{Fixing the values of $f_{Q,u,d}$}
Eqs. (\ref{4Dm},\ref{fs}) provide eight relations, between the flavor
observables, for
the nine flavor parameters, $f_{x^i}\,,$ and the overall scale $\lambda_{5D}\,.$
In order to maintain the perturbativity of the theory
$f_{x^i}$ ($\lambda_{5D}$) cannot be arbitrarily small
(large) as follows.
For the theory to contain at least two KK modes before it
becomes strongly coupled we
require~\cite{cust1}~\footnote{In~\cite{nomura2} 
higher value of $\lambda_{5D}$
  are considered. This implies that the 
the KK modes are strongly coupled.
  Since we will consider this kind of coupling in our analysis below,
  we cannot trust our results in that range. Following the
  spirit of~\cite{nomura2},
however, who argue that in the above limit
  nothing special is expected to happen to observable quantities
  we claim that our conclusions should likewise hold in the above range.}
\beq
\lambda_{5D}k\lsim4\,.\label{lam5D}
\eeq
$f_{x^i}$ cannot be smaller than unity
since it implies that the corresponding bulk mass, $k \; C_{x^i}$, exceeds
the curvature scale so that    
$\Psi_{x^i}$ should be treated as a brane localized fermion.
In addition to have a sufficiently heavy partner for the
$SU(2)_R$ partner of $t_R$ (in order to avoid large 
shift in coupling of $b_L$ to $Z$ via its mass mixing with $b_L$)
requires~\cite{cust1}
\beq
f_{u^3}\gsim1.2\,.\label{fu3}
\eeq
Note that constraint from EWPM ($Z \rightarrow
b \bar{b}$) typically requires~\cite{cust1,nomura2} 
$f_{Q^3}\gsim2$ regardless of flavor mixing. 
%
We will be careful in what follows to correlate (incorporate)
this constraint with (into) our flavor analysis. In the previous
study with few TeV KK masses \cite{Burdman:2003nt}, such correlation was not studied.
Thus, by itself, 
this bound combined with the known value of $m_t$ 
and the lower bound of~(\ref{fu3}) 
effectively
yield a 
lower bound on $\lambda_{5D}\,,$
of ${\cal O}\left(3\over k\right)$. This is close to upper bound of~(\ref{lam5D}),
which is important for our results related to 
radiative B decays (see subsection~\ref{Dipole}).
As for constraint from FCNC processes we shall see below that the
smaller 
$f_{Q^3}$ is the larger are the NP contributions.

Consequently, to derive lower bound on these contributions we shall use
the lower bound of~(\ref{fu3}) 
and upper bound of~(\ref{lam5D}) 
which gives maximum value of $f_{Q^3}$ of ${\cal O}(3)$.
%
%
%

Having fixed the values of $\lambda_{5D}$ and $f_{u^3}$
we can use eqs. (\ref{4Dm}) and (\ref{fs}) to solve for the other eight parameters.
In table \ref{fstab} we summarize the related relation between these parameters
and the resultant values.
At this point, neglecting effects of higher KK fields,
we have all the information regarding the IR limit of
the theory. This is however not very interesting since
apart from the scalar sector, which is model dependent, it is
equivalent to the SM.
In order to study its deviation from the SM we need to consider the
higher KK modes.

\begin{table}[!hbt]\begin{center}
 \begin{tabular}{||c|l|l|l||}
    \hline\hline
    { Flavor}& { $f_Q^{-1}$} & { $f_u^{-1}$} & { $f_d^{-1}$}\cr
    \hline\hline
    I &$ {{ \lambda^{3}}\over { f_{Q^3}}}\sim 0.4 \times
    10^{-2}$&
    ${ m_u \over m_t\,} { f^{-1}_{u^3}\over \lambda^3} \sim10^{-3}$&
     ${ m_d\over m_b}\,{ f^{-1}_{d^3}\over \lambda^3} \sim 10^{-3}$
    \cr \hline
    II&$ {{ \lambda^{2}}\over { f_{Q^3}}}\sim
    2\times 10^{-2}$&
    ${ m_c \over m_t} \,{ f^{-1}_{u^3}\over  \lambda^2} \sim 10^{-1}$&
      ${ m_s\over m_b}\,{ f^{-1}_{d^3}\over \lambda^2} \sim 0.3\times 10^{-2}$
    \cr\hline
    III &$ {f_{u^3} m_t \over v\lambda_{5D}k}\sim{1\over3}$ &${\cal {\mathbf
    O}}\left({\mathbf {5\over6}}\right)$&
    ${{ m_b\over m_t} { f^{-1}_{u^3}}}\sim 0.6\times 10^{-2}$
    \cr
\hline\hline\end{tabular}
\caption{{\small The known quark masses and CKM mixing implies relation
    between the model flavor parameters, $f_{x^i}$,
    (\ref{4Dm},\ref{fs}).
    The value of $f_{u^3},\lambda_{5D}$ is determined by requiring the
    theory is
    perturbative (\ref{lam5D},\ref{fu3}).}}\label{fstab}
  \end{center}
\end{table}

\subsubsection{Flavor violation with KK modes}
In generic split fermions models the flavor structure of the KK
theory is complicated and cannot be calculated analytically.
The bulk RS1 framework, however has several unique feature which
makes, at leading order, the flavor structure of the KK theory
extremely simple.
The above structure is understood from the above observation:
{\it At leading order, all the profiles of the KK modes are localized
  near the TeV brane.}

This implies that flavor violating coupling are ``KK blind''.
In particular, 
$c$'s for all SM fermions (except for $t_R$ which plays a minor
role in low energy flavor dynamics) are close to $1/2$ so that
one finds the following features for the fermion KK excitations:
 \begin{itemize}
 \item[(i)] Up to small corrections (due to difference in widths),
   for a given KK level, 
all the KK excited fermion profiles are the same.
\item[(ii)] The masses of the three generations for each KK level, 
are degenerate, up to a small correction.
  \end{itemize}
This holds to a very good approximation 
as shown in Appendix \ref{appkkfermion} for the fermion ones.
The reason behind this result is that KK
spectrum and wavefunctions are
not sensitive to this (minor) variation in $c$ (in contrast to the
zero-mode wavefunction, which is very 
sensitive to $c$).

Our next step is to explicitly present the part of the KK
Lagrangian yielded after we apply the integral over 
$z$. This part  
describes the flavor structure of the theory including
interactions with the higher
KK modes.
\beq
{\cal L}_{\rm KK}={\cal L}^g_{\rm KK}+{\cal L}^Y_{\rm KK}+{\cal L}^Z
\label{LKK}\,,
\eeq
where ${\cal L}^g_{\rm KK}$ contains the interaction with the higher
KK gauge bosons, ${\cal L}^Y_{\rm KK}$ contains interactions
with higher KK fermions and ${\cal L}^Z$ contains the flavor violating
part due to EWSB effects.
After integrating over 
$z$
the gauge interactions part, in the
``special basis'', is of the form
\beq
{\cal L}^g_{\rm KK} \sim \sum_{x,i}& \sqrt{k\pi r_c}&g_{x}\sum_{n}
G^n\left[\psi^{0\dagger}_{x^i}\psi^0_{x^i}\left({1\over k\pi
      r_c}+{1\over f_{x^i}^2}\right)\right.
\nonumber \\
&&\left.
  + \sum_{m} \left({1\over
      f_{x^i}}\,\psi^{0\dagger}_{x^i}\psi^m_{x^i}+
   \sum_p \psi^{m\dagger}_{x^i}\psi^p_{x^i}+h.c \right)\right]
\label{Lg}\,,
\eeq
where $\psi^l_{Q,u,d}$ is a 4D fermion field, $i$ is a flavor
index, $l,m,n$ stands for 
the corresponding KK levels, $g_x$ stands for the three $4D$/SM gauge
couplings (see Eq. (\ref{0mode})),
$G^n$ is a KK gauge boson
and we suppressed the Lorenz structure. In addition, we neglected EWSB
effects which are separately discussed below.
%
%

The Yukawa part is of the form~\footnote{We shall work in the mass insertion
  approximation. That is we shall treat the Yukawa interactions/mass
  terms on the TeV brane as a perturbation.}
\beq
\hspace*{-.1cm}{\cal L}^Y_{\rm KK}\approx  h \sum_{m,i,j} 2 \lambda_{u,d\,
        5D k}^{ij}
 \left. \hspace*{-.05cm} \left( 
{ \psi^{ 0 \dagger }_{ Q^i } 
\over{ f_{Q^i} } } \,\psi^m_{u^j,d^j}
    + \psi^{m\dagger}_{Q^i} \, { \psi^0_{ u^j,d^j } \over{ f_{u^j,d^j } } }
 + \sum_n
    \psi^{ m \dagger }_{ Q^i } \psi^n_{ u^j,d^j } + h.c \right)
\right|_{\rm TeV} \hspace*{-.2cm}.
\label{LY}
\eeq
%
%

\subsubsection{Flavor violation in $Z$ coupling from EWSB}~\label{Zshift}
In our framework, with or without the Higgs,
EWSB occurs only at the boundaries of the extra dimension.
This leads to an important effect relevant to our considerations.  
That is, it distorts profile of the physical $Z$ near the TeV
brane~\cite{EWPM,cust1, nomura2}.
Its profile is given by a linear combination of the undistorted
KK states; where the mixing angle, $\delta g_Z$, between the ordinary basis and the
distorted one is of order of 
$$\delta g_Z\sim \sqrt{ k\pi r_c } \left({M_Z\over m_{\rm KK}}\right)^2 ={\cal
  O}(1\%)\,.$$

Below we shall only be interested in the flavor violating part of
the $Z$ coupling to two
SM fermions. Thus the relevant part of the Lagrangian for this case
is given by~(\ref{Lg})
\beq
{\cal L}^Z\sim \sum_{x,i} \sqrt{k\pi r_c}\,\delta g_{Z} {g_2\over
  2\cos \theta_{\rm W}}
Z\psi^{0\dagger}_{x^i}(v_f-\gamma_5 a_f)\psi^0_{x^i}\,{1\over f_{x^i}^2}
\label{LZ}\,,
\eeq
where $a_f=T_3^f$ and $v_f=T_3^f-Q_3^f \sin^2\theta_{\rm W}\,.$
%
%

\subsection{Flavor violation and spurion analysis}~\label{FVSA}
Using the values of the flavor parameters in table \ref{fstab}
and the flavor structure of the theory given in
eqs. (\ref{Lg},\ref{LY},\ref{LZ}) the model is fully determined.
At this point, in principle, one can derive a prediction for any
process which is
related to the flavor sector of the theory.
It is very instructive, however, to note that the above
framework has an underlying organized structure.
It is linked with our above observation 
that, to leading order, flavor violation  
appear in a universal way in the KK couplings.

We can summarize the relation in eqs. (\ref{Lg},\ref{LY},\ref{LZ}) by the
following:
\begin{itemize}
\item Flavor violation in coupling between KK modes stems only from the
Yukawa matrices $\lambda_{u,d \, 5D}\,.$
\item Flavor violation between a zero mode and other fields is
  always accompanied by a factor of $f^{-1}_{x^i}\,.$
\end{itemize}
So far all our analysis [in particular
(\ref{Lg},\ref{LY},\ref{LZ})]  was done in the ``special basis''
in which the 5D bulk masses are diagonal.
In order to get more insight into the pattern of the flavor violating
interactions we consider the  global
symmetries of the above KK theory in various limits.

Switching off all the interactions we find the following large flavor
symmetry
 $${\rm U(3)}_{Q,u,d}\times \Pi_n {\rm U(3)}^{n}_{Q,u,d}\,,$$
where the first term stands for the SM flavor group and the second
stands for product of groups, one for each KK level (per representation). 
Omitting, for a moment, the zero modes we find that the 5D Yukawa
matrices break the above symmetries as follows
$${\rm U(3)}_{Q}^n\times{\rm U(3)}_{u,d}^m\, \ 
{\overseta{{{\hspace*{-.54cm}\lambda_{u,d\, 5D}}}}{\longrightarrow}}
\, \
{{\rm U(1)}^3}_{u',d'}\,,\ \ \ \ \ n,m\neq0\,.$$
This implies that we can think of the 5D Yukawa matrices as,  spurion fields,
bi-fundamentals
of the diagonal KK flavor group, ${\rm U(3)}^{n}_{Q}\times{\rm U(3)}^{n}_{u,d}\,,$
$$\lambda_{u,d\,5D}=(\bar {\mathbf 3}_Q^n,{\mathbf3}_{u,d}^n)\,.$$

As discussed above, the only way zero modes can couple to
other fields (we are not interested in the flavor universal
pieces) is through extra factors of $f_{x^i}^{-1}\,.$
Thus from eqs. (\ref{Lg},\ref{LY},\ref{LZ}) we find the 
following breaking pattern
$${\rm U(3)}_{Q,u,d}\times{\rm U(3)}_{Q,u,d}^n\, \ 
{\overseta{{{\hspace*{-.64cm}f^{-1}_{Q^i,u^i,d^i}}}}{\longrightarrow}}
\, \
{{\rm U^D(1)}^3}_{Q,u,d}\,,$$
where ${\rm U(3)}_{Q,u,d}^n$ stands again for the diagonal KK flavor
group and U$^{\rm D}$(1) is in the diagonal basis of 
the KK and SM flavor groups.
This implies that we can view the $f_{x^i}^{-1}$ as 
eigenvalues of a matrix, we denote as $F_x$, where in the ``special
basis'' we have
$$F_x
\equiv{\rm diag\,}\left(f_{x^i}^{-1}\right)\,.$$
We learn then that $F_x$ transforms as a bi-fundamental under
the~${\rm U}(3)_{Q,u,d}\times~{{\rm U}(3)^{n}_{Q,u,d}}$\\ flavor group
$$F_Q^\dagger,F_{u,d}=(\bar {\mathbf
  3}_{Q,u,d}^n,{\mathbf3}_{Q,u,d})\,.$$

We finish this part by noting that there 
are two interesting limits regarding the spurions $F_{Q,u,d}\,.$
\begin{itemize}
\item[(i)]
$F_{x^i}\to0$ -
the SM flavor group is unbroken. Looking at the values of 
$f_{x^i}^{-1}$ given in table \ref{fstab} we find the following
feature. All the values of the $f_{x^i}^{-1}$ apart from the ones related
to the top mass are small.
This implies that the model has a built-in approximate flavor
symmetry for the light quarks.
This is indeed the reason why the framework may avoid
the severe constraint from FCNC processes with such a low KK masses.
We can compare this with the flat extra dimension models which
require KK masses of ${\cal O}({\rm 1000\,TeV})\,.$
\item[(ii)] $F_{x}\to constant\times {\mathbf 1}_3$ - 
the ${\rm U(3)}_{Q,u,d}\times~{\rm U(3)}^{n}_{Q,u,d}$ 
flavor group is broken to a $U(3)$ diagonal one.
This means that to have flavor violation (not through SM Yukawa interactions) 
a non-degeneracy in $F_{Q,u,d}$ is required.
This is the RS-GIM mechanism and 
as discussed above
only top related entries in $F_{Q,u}$ induce {\em sizable} RS-GIM
violation.   
RS-GIM mechanism
is violated by 1st and 2nd generation as well in our framework
since $F_{Q, \; d, \; u}$'s are non-degenerate. However, 
$F_{Q, \; d, \; u}$'s are small so that
violation of RS-GIM is not severe, i.e., FCNC's are protected by built-in
approximate symmetries.
Also, the 2nd limit corresponds to minimal flavor violation (MFV)
since only source of FV is $5D$ Yukawa which
is the same spurion as the 
$4D$ Yukawa since $F \propto {\mathbf 1}_3$.
\end{itemize}

\subsubsection{Relations among couplings and
  non-trivial predictions}\label{relation}

In the KK theory the number of vertices with nontrivial
flavor structure is large.
According to our approximation that flavor violation is KK blind, 
these interactions are described by only
five spurions $F_{Q,u,d}$ and $\lambda_{u,d\, 5D}$.
Consequently these couplings are not independent and there
are relations among them.
One such trivial relation is that
the Yukawa interactions between any two KK fermions, with KK levels 

$n,m$, (and the Higgs), $\lambda_{u,d}^{nm}$,
are just proportional to a single spurion $\lambda_{u,d\, 5D}\,,$
$$\lambda_{u,d}^{nm}\propto\lambda_{u,d\, 5D}\,.$$
The ones that are relevant to our work contain at
least a single zero mode leg. These have less trivial relations
among them.
For example the product of a gauge interaction between a zero mode and
an n KK fermion (and a KK gluon) and a Yukawa coupling between n and
m KK fermions (and the Higgs) is proportional to the direct Yukawa coupling between a
zero mode and an m level KK fermion (\ref{Lg},\ref{LY}):
\begin{equation}
  g^{0n}_{Q} \lambda_{u,d}^{nm}\,,\,  
\lambda_{u,d}^{nm} g^{m0}_{u,d} \,\, \propto\, \, gF_{Q}
  \lambda_{u,d\, 5D }\,,\, g\lambda_{u,d\, 5D} F_{u,d}\,\, \propto\, \, 
\lambda_{u,d}^{0m}\,,\,\lambda_{u,d}^{m0}
 \label{relS} \end{equation}

We can use the above to argue that the
KK gluon diagram shown in fig. 4. [discussed below in 
subsection (\ref{Dipole})] yields a small
flavor-violating effect since it is aligned with the down type mass matrix
\begin{equation}
  g^{0 m}_Q \lambda_{d\, 5D}
g^{n 0}_d\propto \lambda_{d\, 4D}\,. \label{align}
\end{equation}
This basically explains why the lowest order,  sizable, contribution to
chirality flipping operators that we find in subsection (\ref{Dipole})
is proportional to
${\cal O}\left(\lambda_{u,d\, 5D}^3\right)\,.$ 

We finish this part by pointing out that the above relations 
yield remarkable
correlation between measurements of observables in low energy
experiments and ones related to high energy theory specific to 
this framework. In principle just based on low energy experiments (and
top mass measurement) we can determine all the model flavor parameters
{\it i.e.} the values of $f_{x^i}$, the mixing angle and the CP phases.
Consequently, using relations similar to the ones in
eqs. (\ref{relS},\ref{align}), we will (in principle) be able 
to completely predict the amplitude for high energy processes in which
incoming or outgoing, on shell, KK quarks are participating!

\subsubsection{Counting parameters in the KK theory}\label{SpIm}
The above analysis shows yet another special feature of the RS1
framework.
That is we can directly translate the flavor violating parameters
in the 5D theory to the ones appearing in the couplings of the KK theory.
In the ``special basis'' this is transparent since there is one to one
correspondence
between the eigenvalues of $C_{Q,u,d}$ (and $\lambda_{u,d\,5D}$) and
$F_{Q,u,d}$ (and $\lambda_{u,d\,5D}$)
which 
are the flavor violating sources in the 5D and the KK theories
respectively.
Note that in flat extra dimension models there is no such simple
correspondence since flavor violation in the KK theory is found by
calculating overlap-integrals between the WF's of the fields.

Let us verify that statement by counting the flavor parameters in
the KK theory and see that we can reproduce our results derived in the 5D
theory summarized in tables \ref{down} and \ref{Tot}.

As done in subsection~\ref{FV5D},  we start with the
case in which flavor violations occur only in
  the down type quark sector, $F_{u},\lambda_{u\, 5D}\to0\,.$
  Then we consider the generic case.
  \begin{itemize}
  \item[(i)] Down type quark sector - Flavor violation is encoded in
    three generic $3\times3$ matrices, $F_{Q,d}$ and
    $\lambda_{d\, 5D}$ which contain altogether 27 real and 27
    imaginary parameters. We can use the diagonal KK and the SM flavor
    symmetries,  
    ${\rm U(3)}_{Q,d}^n\times{\rm U(3)}_{Q,d}\,,$ to eliminate
    $4\times 6-1=23$  phases [there is still a
    conserved U(1) baryon symmetry in the full theory].
    Thus altogether we find four physical phases as in
    table~\ref{down}.
    Two of these are CKM like phase in $D_{L,R}$ and the
    other two are ``Majorana-like", flavor diagonal, and can be shifted between
    $D_{L,R}\,,$ (for more details see appendix~\ref{Ddn}).
     Similarly we can remove $4\times3=12$ real parameters out of the
     27 ones and hence 15 physical real parameters are left as in table~\ref{down}.
    The real parameters are three quark masses, six mixing angles
    related to $D_{L,R}$ and the six eigenvalues of  $F_{Q,d}$ which
    measure the non-universal couplings between the different
    generations and the KK gauge fields.
  \item[(ii)] Generic model -  Flavor violation is encoded in
    five generic $3\times3$ matrices, $F_{Q,u,d}$ and
    $\lambda_{u,d\, 5D}$ which contain 45 real and 45
    imaginary parameters. We can use the diagonal KK and the SM flavor
    symmetries,  
    ${\rm U(3)}_{Q,u,d}^n\times{\rm U(3)}_{Q,u,d}\,,$ to eliminate
    $6\times 6-1=35$  phases [there is still a
    conserved U(1) baryon symmetry in the full theory].
    Thus we find ten physical phases as in
    table~\ref{Tot}.
    Four of these are CKM like phases in
    $U_{L,R},D_{L,R}$ and the
    other six are `` Majorana-like" and can be shifted between
    $D_{L,R},U_{L,R}\,,$ 
(for more details see appendix~\ref{Ddn}).~\footnote{Note that we treat 
$V_{\rm CKM} = U_L^\dagger D_L$ 
and also the analogue matrix which appear in the RH charged currents, 
$V_{\rm CKM}^R = U_R^\dagger D_R$, as dependent
matrices to avoid double-counting of phases.}
Similarly we can remove $6\times3=18$ real parameters out of the
     45 ones and hence 27 physical are left as in table~\ref{Tot}.
    The real parameters are six quark masses, twelve mixing angels
    related to $U_{L,R},D_{L,R}$ and the nine eigenvalues of  $F_{Q,u,d}\,$.
 \end{itemize}

 The last point that we want to make here is related to flavor
 diagonal CPV.
 We demonstrate that unlike, for example in generic SUSY models, 
this
framework does not contain flavor diagonal phases in
the sense that, with{\em out} flavor mixing,
the Majorana phases mentioned above do not contribute to EDMs, i.e., are not physical.
This can be shown by considering  the limit in 
which flavor violation is absent. 
In that case, all the
 ten CPV phases can be removed and are not physical:
No flavor violation implies that 
 the spurions $\lambda_{u,d\, 5D}$
 and $F_{Q,u,d}$ (or $C_{Q,u,d}$) can be diagonalized simultaneously.
Then one can use a chiral rotation to remove the phases
in the Yukawa matrices and to eliminate all the phases from the theory 
(apart from the strong CP phase).

\section{CFT interpretation of flavor structure of RS1}
\label{cft}

In this section, we will show that, 
remarkably, there is an understanding of flavor structure/built-in approximate
symmetries in CFT picture as well. 
The dual description of 
this RS1 model has been discussed before (see, for example,
references \cite{dual, cust1, Contino:2004vy} in addition to
\cite{rscft}). For completeness,
we will review this description and then describe 
the flavor structure in CFT picture which has not been discussed in detail before. 

As per AdS/CFT
correspondence, RS1 is dual to a strongly coupled CFT of which the minimal Higgs is a
composite arising after conformal invariance is broken.
The SM gauge and fermions fields originate as fundamental fields/external to CFT, but
coupled to the CFT/Higgs sector.
Due to this coupling,
these external fields mix with CFT composites, 
the resultant massless states correspond to the 
SM gauge and fermion fields (these are dual to the {\em zero}-modes
on RS1 side). 
The degree of this mixing depends on the anomalous/scaling dimension
of the CFT operator they couple to. The coupling of SM gauge bosons and fermions 
to Higgs goes via their composite
component since
Higgs is a 
composite of the CFT.
Thus, this coupling of fundamental gauge and fermion fields to 
CFT operators is essential for gauge boson and fermion masses to arise at 
the weak scale.


\subsection{Duality at qualitative level}
\label{cftqual}

We begin with a qualitative description of dual CFT.
The dual interpretation of gauge 
fields in bulk
is that the $4D$ CFT has a conserved global symmetry current
(which is a marginal operator, i.e., zero anomalous dimension) coupled to 
a $4D$ gauge field: $A_{ \mu } J^{ \mu }_{CFT}$. This is
just like the photon coupling to a $U(1)_{em}$ global symmetry current of QCD.

The operator $J^{ \mu }_{CFT}$ interpolates/creates
out of the vacuum
massive spin-$1$ composites of CFT 
(``techni-$\rho$'s''
in case of, global, electroweak symmetry of CFT).
These composites are similar to $\rho$-mesons in real QCD and 
are dual to
gauge KK modes on RS1 side.

Similarly,
the dual interpretation of 
a bulk fermion is that there is a fundamental fermion 
(external to CFT) coupled to fermionic CFT operator: $\psi {\cal O}_{CFT}$.
The operator ${\cal O}_{ CFT }$ interpolates/creates out of vacuum 
masssive spin-$1/2$ composites
(just like $J^{ \mu }_{ CFT }$ creates spin-$1$ composites) which are dual to
fermion KK modes on RS1 side.

The
$c$ parameters (bulk fermion masses) are dual to scaling dimension of ${\cal O}_{CFT}$'s
which control the mixing of fundamental fermions with CFT composites.
The choice of $c > 1/2$ for light fermions is dual to 
a coupling
$\psi {\cal O}_{CFT}$ being irrelevant so that the mixing 
of $\psi$ with composites is small, i.e., the corresponding SM fermion is mostly fundamental.
Thus, the coupling of SM fermion to 
composite Higgs and also to spin-$1$ composites 
is small since both couplings have to go via the small mixing: the
small coupling to $\rho$'s suppresses FCNC's from their tree-level
exchange\footnote{This is 
the flavor-dependent part of the coupling. There's also
a universal coupling induced by $\gamma - \rho$ mixing.} \cite{cust1}. This agrees qualitatively
with small $4D$ Yukawa and small flavor-dependence in the coupling to gauge KK mode
obtained on $5D$ side. Thus, it is easy to see how the notions of
approximates symmetries and RS-GIM arise in the CFT picture.

Similarly one can see the tension arise for the 
third generation as follows. The
SM top quark should have large composite component
so that it has $O(1)$ coupling to the composite Higgs,
i.e., {\em fundamental} top quark should have relevant coupling to CFT
(dual to $c < 1/2$).\footnote{$c = 1/2$
corresponds to {\em marginal} coupling just like $A_{ \mu } J^{ \mu }_{CFT}$.}
However, if the fundamental $(t, b)_L$ has relevant couplings to the
CFT sector, then SM $b_L$ will
have large couplings to $\rho$-mesons (due to large mixing of fundamental
$b_L$ with composites) leading to a shift in coupling of
SM $b_L$ to $Z$.
So, $Z \rightarrow b \bar{b}$ requires that coupling of fundamental $(t, b)_L$
to CFT be at most {\em mildly} relevant (dual to $c \sim 0.3-0.4$). Nonetheless,
the coupling of
SM $b_L$ to $\rho$-mesons is still larger than that of light fermions
and there {\em is} a small shift in coupling of SM $b_L$ to $Z$, 
leading to FCNC's discussed earlier. 

Also, to obtain $\lambda_t \sim 1$ with only mildly relevant coupling
of $t_L$ requires that the coupling of the fundamental $t_R$ to the
CFT sector must be 
more relevant, which is dual to $c$ for $t_R \lsim0$. Thus, 
SM $t_R$ contains a sizable admixture of composites.

We see that particles localized near the TeV brane ($t_R$ zero-mode, 
Higgs, {\em all} KK modes) are (mostly) composites
in the CFT picture. This is expected since 
TeV brane corresponds to the IR of the CFT. Thus
particles which are localized near that brane 
correspond to IR degrees of freedom in (i.e., composites of) CFT.
Similarly,
particles localized near Planck brane (light fermion zero-modes) 
are (mostly) fundamental/external in  
the CFT picture. This is expected since 
Planck brane corresponds to the UV of the CFT so that
particles 
localized near that brane correspond to UV degrees of freedom in the CFT
picture.

\subsection{Duality at semi-quantitative level}
\label{cftquant}

So far, the CFT description 
(including the dual understanding of flavor structure/built-in
approximate symmetries
of the RS model) was qualitative.
In this section, we will obtain a
semi-quantitative understanding of flavor structure/RS-GIM,
in particular, Eqs. (\ref{Lg}) and (\ref{LY})
in the CFT description.
For this purpose, we
assume that the CFT is like 
a large-$N$ ``QCD'', i.e., SU(N) gauge theory 
 with some ``quarks''.

Before considering couplings of fermions, as a warm-up, we begin 
with coupling of Higgs to gauge KK mode
(see, for example, \cite{dual}). On $5D$ side, 
this coupling
$\approx g \sqrt{ 2 k \pi r_c }
\approx \sqrt{ 2 g^2_{ 5D } k }$
[see Eqs. (\ref{gaugeKKHiggs}) and (\ref{0mode})]:
for simplicity, we omit the three SM gauge groups indices
in the following).
Since these are three
particles all of which are localized near TeV brane, 
in CFT picture, this is a coupling of $3$ composites.
We will use the result of a large-$N$ QCD theory in which naive dimensional analysis
(NDA) estimation yields:
\begin{equation}
\hbox{coupling of} \; 3 \; \hbox{composites} \sim \frac{ 4 \pi }{ \sqrt{N} }
\end{equation}
(see, for example, reference \cite{witten}). 
With a coupling of this size, 
loops are suppressed
by $\sim 1/N$ compared to tree-level. 

Assuming duality, we equate the above two couplings to obtain 
the following relation between $N$ (number of colors of CFT)
and parameters of the $5D$ theory
\begin{equation}
\sqrt{ g_{ 5D }^2 k } \sim \frac{ 4 \pi }{ \sqrt{N} }
\label{g5DN}
\end{equation}
Is there a consistency check of this relation? 
The answer is yes by comparing
low-energy gauge coupling on the two sides 
(see 6th reference of \cite{rscft}).
On the CFT side, we get
$$1/ g^2 \sim N / \left( 16 \pi^2 \right) 
\; \log \left( k / \hbox{TeV} \right)\,.$$
This is due to contributions of CFT quarks to running of external gauge couplings
from the Planck scale down to the TeV scale (just like contribution of
SM quarks to running of $\alpha_{QED}$);   
whereas, using $\log \left( k / \hbox{TeV} \right)
\sim k \pi r_c$,
we can rewrite the {\em zero}-mode (i.e., low energy)
gauge coupling on $5D$ side (see Eq. (\ref{0mode})) as
\beq
1/g^2 = \log \left( k / \hbox{TeV} \right) / \left( g_{ 5D } ^2 k \right)\eeq
These two gauge couplings agree using the relation in Eq. (\ref{g5DN})
\footnote{Here, we have assumed 
that the gauge coupling in the CFT picture has 
a Landau pole at the Planck scale -- 
this is dual to small Planck brane kinetic terms 
on RS1 side (see 6th reference of \cite{rscft} and \cite{dual}).}.
In particular, we see that $N \sim 5-10$ is required to get $O(1)$ low-energy
gauge coupling.

We now move on to couplings of fermions
which will give us a semi-quantitaitive understanding of flavor structure
using the CFT picture.
Begin with couplings of fermions to Higgs. 
The coupling of $2$ KK fermions to Higgs
is $\approx 2 \lambda_{ 5D } k$ and
replacing KK mode (localized near TeV-brane) 
in this coupling
by zero-mode fermion we get a suppression in $5D$ picture
(due to wavefunction at TeV brane of zero-mode vs. KK mode) 
of $\sim 1 / f_{x^i}$
(see Eqs. (\ref{lambda4D}) and (\ref{LY})).

In the CFT picture, the coupling of $2$ KK fermions and Higgs 
(again, three particles localized near TeV brane) is a coupling of
$3$ composites. Also, coupling of SM/physical 
fermion to composite Higgs 
must involve its composite component, i.e., we have to pay the price of
mixing $\equiv \xi_{x^i}$ between fundamental fermion and CFT composite
each time we
replace a KK fermion by a zero-mode/SM fermion in the above coupling.
So, we get (for example, for $Q$ and $d$ modes)
\begin{eqnarray}
{\cal L}^Y_{ 0, \; KK } & \sim &
h \frac{ 4 \pi } { \sqrt{N} } \left( \psi_Q^m \psi_d^n + \xi_Q \psi_Q^0 \psi_d^n
+ \xi_{ d } \psi_d^0 \psi_Q^n + \xi_Q \xi_d \psi_Q^0 \psi_d^0 \right)
\label{LYdual}
\end{eqnarray}
Assuming duality, we equate Eq. (\ref{LYdual}) and Eqs. (\ref{lambda4D}) 
and (\ref{LY}) to obtain\footnote{Since, in the CFT picture, 
mixing depends on anomalous dimension of fermionic operator and, on RS side, $f_x^i$ 
depends on $c$
parameter, 
using Eq. (\ref{mixingf}), we obtain
a relation between $c$ parameter (i.e., $5D$ mass of fermion)
and the anomalous dimension of fermionic operator
which agrees with the standard AdS/CFT dictionary (see, for example,
\cite{Contino:2004vy}).
}  
\begin{equation}
2 \lambda_{ 5D } k \sim \frac{ 4 \pi }{ \sqrt{N} }
\end{equation}
and
\begin{equation}
\xi_{ Q, d } \sim \frac{1}{ f_{ Q, d } }
\label{mixingf}
\end{equation}
Using $N \sim 5-10$ (obtained before), we get
$2 \lambda_{ 5D } k \sim 5$
-- this size of $\lambda_{ 5D } k$ agrees with the one before (based on top Yukawa).

We can check the relation in Eq. (\ref{mixingf}) using coupling of fermion to gauge KK mode. 
The coupling of gauge KK mode to $2$ {\em KK} fermions
(again, three particles localized 
near TeV brane) is similar to 
gauge KK coupling to Higgs, i.e., in the CFT picture, it is a  
coupling of $3$ composites.
As in the case of coupling to Higgs, replacing a KK fermion by 
zero-mode/physical fermion in this coupling
costs $\xi_x^i$ in the CFT picture
so that we get (for simplicity, we show couplings of $Q$ modes only):
\begin{eqnarray}
{\cal L}^g_{ KK \; composite } & \sim & 
\frac{ 4 \pi } { \sqrt{N} } G^n \left( \psi_Q^m \psi_Q^n + \xi_Q \psi_Q^0 \psi_Q^n
+ \xi^2_Q \psi_Q^0 \psi_Q^0 \right) 
\label{Lgdual1}
\end{eqnarray}
These couplings in CFT picture agree 
with Eq. (\ref{Lg}) using Eqs. (\ref{g5DN}) and (\ref{mixingf}).

In Eq. (\ref{Lgdual1}), we considered
the coupling
involving {\em composite} component of gauge
KK mode so that we had to use the composite component of
zero-mode fermions as well which cost $\xi$ (KK fermions are mostly composite)
and hence the subscript ``composite'' in Eq. (\ref{Lgdual1}). 
However, the
gauge KK mode also has an
elementary component 
since the elementary gauge field mixes with spin-$1$ composites ($\rho$-mesons).
It turns out that 
gauge field is like a fermion
with $c = 1/2$
(for example, zero-mode of gauge field has flat profile just like
a fermion with $c = 1/2$)
so that mixing of elementary gauge boson with $\rho$-meson is given by
(using Eq. (\ref{mixingf}) and Eq. (\ref{fapp}))
\begin{equation}
\xi_{ gauge } \sim \frac{1}{ \sqrt{ k \pi r_c } }
\end{equation}
Then, the
elementary component of gauge KK mode gives the following couplings.
Here, we have to use elementary component of KK
fermion which costs $\sim \xi_x^i$ (zero-mode fermion
is mostly elementary):
\begin{eqnarray}
{\cal L}^g_{ KK \; elementary } & \sim & g \xi_{ gauge } G^n \left( \xi_Q^2 
\psi_Q^m \psi_Q^n + \xi_Q \psi_Q^0 \psi_Q^n + \psi_Q^0 \psi_Q^0 
\right),
\label{Lgdual2}
\end{eqnarray}
where $g$ is coupling of elementary gauge boson.
The
last coupling in Eq. (\ref{Lgdual2})
agrees with flavor-independent coupling in Eq. 
(\ref{Lg}) (using Eq. (\ref{g5DN})), whereas 1st and 2nd couplings
in Eq. (\ref{Lgdual2}) are too small and hence were not shown
in Eq. (\ref{Lg}).

To summarize, in the CFT picture
the factor of $1/f$ each time a zero-mode couples to gauge/fermion 
KK modes or Higgs
(apart from universal coupling to gauge KK mode) is 
due to mixing of
fundamental and composite fermions. This mixing is
required in order for physical fermion (which
is the resultant of this mixing) to
couple
to composites of CFT (i.e., KK modes/Higgs).
Thus, we see that even semi-quantitatively, the small flavor-dependence
in coupling to gauge KK modes is correlated with small Yukawa coupling to Higgs.

\section{Signals}\label{sig}

In the previous parts we focused on studying the
general structure of the flavor sector of our framework.
We found that there is an organizing principle
that allows for a transparent understanding of the structure of
the flavor violating interactions.
We are now at a point at which we can discuss the
phenomenological implications of the above analysis.

Before going into the details we
shall anticipate which class of FCNC processes might be
sensitive to NP contributions.
With $m_{\rm KK}\sim 3\,$TeV
and the approximate flavor symmetries for the light quarks (see subsection~\ref{FVSA})
NP contributions cannot, in general, compete with SM tree level ones.
The same conclusion holds for processes which in the SM are mediated
by QCD penguin diagrams, {\it e.g.} $B\to \phi K_S\,$ as briefly
discussed in~\ref{DF1}.
This is related to the fact that flavor diagonal
couplings between light fermions and
a KK gluon is given by ${g\over\sqrt{k\pi r_c}}$ (see Eq. (\ref{Lg})) 
so that it is
suppressed by $O(5)$ compared with naive 
expectation.

Consequently we shall focus below on three classes of FCNC processes which receive sizable
contributions in the presence of low KK masses, $m_{\rm KK}\lsim 3\,$TeV. 
$\Delta F=2$ processes induced by KK gluon exchange,
$\Delta F=1$ processes induced by a shift in the $Z$ couplings and
radiative B decays which are enhanced due to 
large $5D$ Yukawa (required to obtain $m_t$)
combined with
large mixing angles in the
right handed down type rotation matrix $D_R$.
Finally, we shall discuss the model predictions related to EDMs which are
sensitive to flavor diagonal CP phases.

\subsection{$\Delta F = 2$ processes and the ``coincidence problem''}\label{DF2}

We start by considering the class of $\Delta F=2$ FCNC processes.
These are
mediated through tree exchange of KK gluon as shown in figure 1.
The contributions were already considered in~\cite{huber} but it was
done
for the case with the little hierarchy, {\it i.e.}, with 
$m_{\rm KK} \gsim10\,$TeV which suppresses FCNC 
(in addition to the built-in approximate symmetries and the
RS-GIM mechanism). Consequently, no large effect was found.

Given the couplings between the zero modes and the KK gluons (\ref{Lg})
it is straightforward to estimate the size of the NP contribution.
In terms of spurions the leading NP contribution to 
the $B^0-\bar B^0$ mixing amplitude, $M_{12}^{\rm RS}$,
in the mass basis is proportional to 
\beq
{M_{12}^{\rm RS}}&\propto& 
\left[\left( F_Q F_Q^\dagger \right)_{13}\right]^2\approx
\left[\sum_i\left(D_{L}^\dagger\right)_{1i} f_{Q^i}^{-2} \left(D_{L}\right)_{i3} 
\right]^2 \nonumber\\
&\sim& \left[\left(D_{L}^\dagger\right)_{13} f_{Q^3}^{-2}
  \left(D_{L}\right)_{33}\right]^2
  \sim C_B \left|V_{tb}^* V_{td}\right|^2 f_{Q^3}^{-4} 
\label{DelF2}
\eeq
where $D_{L}$ is a rotation matrix of the down type, left handed, quarks
and $C_B$ is an order one complex number.
Similar contributions proportional to 
$f_{Q^3}^{-2}f_{d^3}^{-2}$ and $f_{d^3}^{-4}$ are subleading 
due to the smallness of $f_d^3$ (see table \ref{fstab}.) and are
therefore omitted above.
We find that magnitude-wise the suppression due to flavor violation
is similar to the SM case.
To estimate the size of the NP contribution we present
its value
normalized by the SM one~\footnote{We use $m_{\rm KK}\sim 3\,$TeV as
  favored by
  the Higgs models. The Higgsless models favors  
smaller KK masses
which will further enhance the NP contributions.}
\beq
{{M_{12}^{\rm RS}}\over M_{12}^{\rm SM}}
     \sim
     {16\pi^2\over N_c}\, {8g_s^2\over g_2^4 S_0(m_t)} {M_W^2\over m_{\rm KK}^2} {k\pi r_c\over
       {f_{Q^3}^4}}\sim{0.5}\times\left({3 {\rm TeV}\over m_{\rm
           KK}}\right)^2
     \left({3\over { f_{Q^3}}}\right)^4 \,,\label{DeltaF2}
\eeq
where $1/N_c=3$ suppression stems from the contraction of the two octet
operators from the two gluonic vertices and $ S_0(m_t)\sim 2.5$ comes
from computing the SM box diagram~(See {\it e.g.}~\cite{Bur} and
refs. therein).

From eq. (\ref{DeltaF2}) we learn that with $m_{\rm KK}\lsim3\,$TeV
even with $f_{Q^3}$
near to its maximal value the NP contributions to $\Delta F=2$ processes are of the same size
as the SM ones.
Furthermore since the above NP contributions come with an arbitrary phase,
[appears in $\left(D_L\right)^*_{33}\left(D_L\right)_{31}$] we expect
also an order one contributions to processes such as, $S_{B\to\psi K_S}$ and
$S_{B\to\rho\rho}$,
the CP asymmetries in $B\to\psi K_S,\rho\rho$~\footnote{for more
  information on $S_{B\to\rho\rho}$ see~\cite{Brho} and refs. therein.},
which, in the SM, 
measures the value of $\sin(2\beta)$ and $\sin(2\alpha)$ respectively.
In addition, a similar derivation yields also sizable contributions to
the imaginary part of $\Delta S=2$ processes,
This implies that
$\varepsilon_K$ also receives NP contributions comparable with the SM
ones
\beq
{\varepsilon_K^{\rm RS}}&\propto& 
Im\left[\left( F_Q F_Q^\dagger \right)_{12}\right]^2\approx
Im\left[\sum_i\left(D_{L}^\dagger\right)_{1i} f_{Q^i}^{-2} \left(D_{L}\right)_{i2} 
\right]^2 \nonumber\\
&\sim& Im\left[\left(D_{L}^\dagger\right)_{13} f_{Q^3}^{-2}
  \left(D_{L}\right)_{23}+\left(D_{L}^\dagger\right)_{12} f_{Q^2}^{-2}
  \left(D_{L}\right)_{22}\right]^2\nonumber\\
  &\sim&  C_\varepsilon \left|V_{td}^* V_{ts}\right|^2 f_{Q^3}^{-4} 
\label{eps}
\eeq
where $C_\varepsilon$ is an order one parameter and
note that both of the contributions from $f_{Q^2}^{-2}$ and from
$f_{Q^3}^{-2}$ are of similar size.
It is clear that~(\ref{DeltaF2}) also holds in this case
\beq
{\varepsilon_K^{\rm RS}\over \varepsilon_K^{\rm SM}}
     \sim {0.5}\times\left({3 {\rm TeV}\over m_{\rm
           KK}}\right)^2
     \left({3\over { f_{Q^3}}}\right)^4 \,,\label{epsK}
\eeq
Finally similar results are also obtained for the NP contributions
related to
$\Delta m_s$, the mass difference between
$B^0_s$ and $\bar {B^0_s}\,.$ 

Consequently, 
the framework predicts sizable CP asymmetry in (e.g.) 

$B_s\to\psi\phi$
\beq
S_{B\to\psi\phi}\sim 1\times \left({2 {\rm TeV}\over m_{\rm KK}}\right)^2
     \left({3\over { f_{Q^3}}}\right)^4\,,\label{psiphi}
\eeq
where the SM prediction is $S_{B\to\psi\phi}\sim{\cal O}\left(\lambda^2\right)\,.$

%
Before studying the implications of these NP contributions,
we point out that in reference \cite{Burdman:2003nt}, smaller
values of $f_{ Q^3}$ were considered such that the constraint from
$Z \rightarrow b \bar{b}$ is not satisfied. This leads to
larger effect in $\Delta F = 2$ processes. In particular,
to suppress the NP contribution to $\Delta m_{ B_d }$ requires
$\left( D_L \right)_{ 13 } \ll V_{ td }$
which is possible only if there are hierarchies in the $5D$ Yukawa, a
possibility that we are not entertaining in this work.

\begin{center}
\begin{eqnarray}\nonumber
\hspace*{3cm}\parbox{150mm}{
\begin{fmfgraph*}(150,150)
\fmfleft{p1,p2}
\fmfright{p3,p4}
\fmf{fermion,label=$d$}{p1,v1}
\fmf{fermion,label=$\bar b$}{v1,p2}
\fmf{gluon,label=$\vspace*{.2cm}G^{(n)^{ }}\vspace*{-.2cm}$}{v1,v2}
\fmf{fermion,label=$b$}{v2,p3}
\fmf{fermion,label=$\bar d$}{p4,v2}
\fmflabel{$F_Q F_Q^\dagger$}{v1}
\fmflabel{$F_Q F_Q^\dagger$}{v2}
\end{fmfgraph*} }
\end{eqnarray}
{\small Fig. 1: Contributions to $\Delta F=2$ processes from KK gluon exchange.}
\end{center}

\subsubsection{``Coincidence'' problem}

Within the SM, experimental data related to the above observables is
 translated into constraints on, 
$\rho$ and $\eta$, the less constrained Wolfenstein parameters. 
The other two parameters, $A$ and $\lambda$ are known to a good accuracy from various
SM tree level processes which are insensitive to our NP contributions.
The fact that the SM can successfully fit, within errors, 
five independent measurements of $\rho,\eta$
supports the SM CKM paradigm~\cite{ckm, ANS_CKM, Nir1}.
Within our framework, however, this SM successful fit
is a pure coincidence! This is since
the above processes receive uncorrelated sizable NP contributions.
Thus, generically, it is not expected that all of them can be fitted
together by only two parameters.

In fig. 2. we show the present SM fit yielded by $\Delta F=2$ processes
in the $\rho-\eta$ plane~\cite{CKM,CKMpresent}.
Due to hadronic uncertainties the ``coincidence problem" is yet not a
severe one~\cite{ENP}.
In the near future, assuming that the SM fit
will continue to be a
successful one, when various uncertainties 
are expected to be brought down and more
measurements will be made the problem will be sharpened.
This is illustrated by fig. 3. which shows the
$\rho-\eta$ plane in the presence of various new constraints 
(assuming NP
contributions are negligible)
from processes which
may become feasible to experiments in the future~\cite{CKMfitter}.

%
The interesting aspect of this coincidence problem is that it leads to
signals!
Since the natural size of NP contribution to $\Delta F = 2$ processes
is comparable to SM, it is clear
that the fit to data in this NP model requires
$\rho, \eta$ which are $O(1)$ different than in the SM fit. 
This implies that the angle $\gamma$ in this model
is also different than in the SM fit. Thus, CP asymmetries
in $B \rightarrow \rho \rho, DK$ which are measurements of $\gamma$
(after subtracting the mixing phase using
$B \rightarrow J / \psi K_s$)
even in the presence of NP (since NP contribution to the decay
amplitude is very small)
will 
deviate by $O(1)$ from SM expectations.
The preliminary measurements of $B \rightarrow \rho \rho, DK$
seem to agree with SM expectation \cite{Charles:2004jd, Poluektov:2004mf} 
and thus
constrain NP contributions of this size,
but the experimental errors are still large.

\vspace*{0.5cm}
\raisebox{0.cm}{\includegraphics[width=8cm,height=8cm]{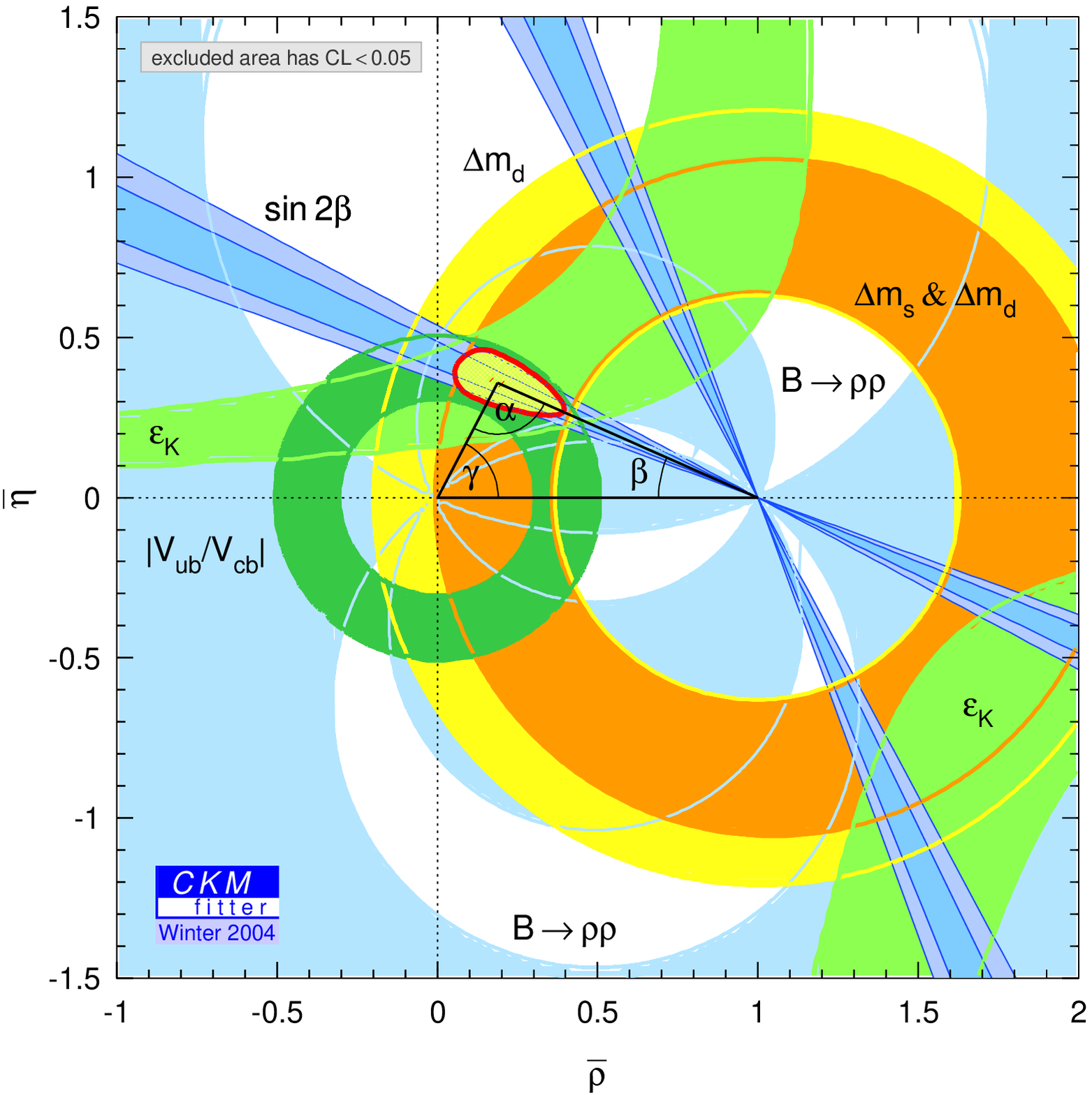}}
\hspace*{.2cm}\\
\vspace*{.1cm}
{\small Fig. 2:
Current constraints on the unitary triangle~\cite{CKMpresent}.}

\vspace*{0.5cm}
\raisebox{0.cm}{\includegraphics[width=8cm,height=8cm]{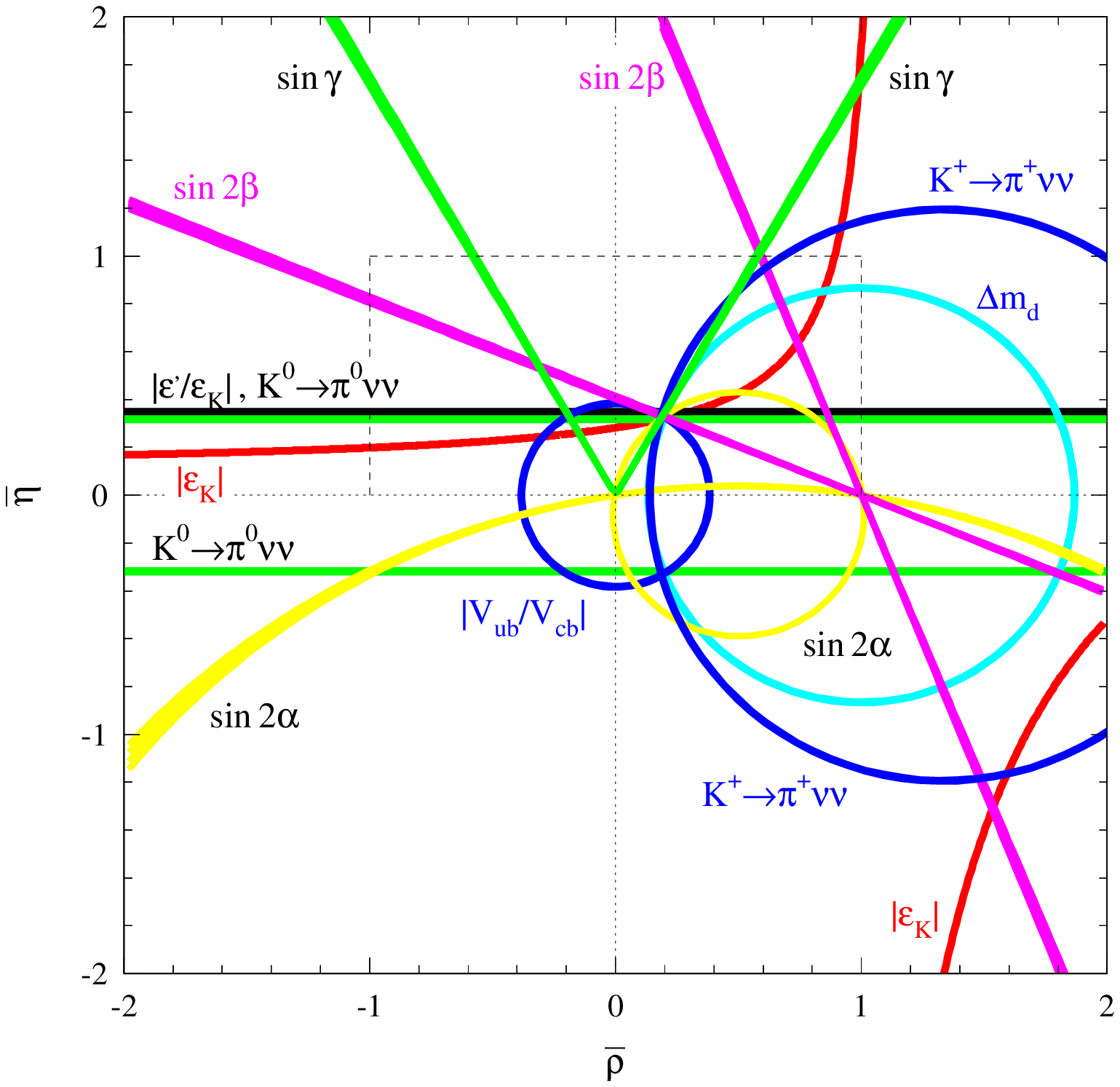}}
\hspace*{.2cm}\\
\vspace*{.1cm}
{\small Fig. 3:
  Illustration: future constraints on the unitary triangle, assuming
  SM~\cite{CKMfitter}.}

\subsection{$\Delta F = 1$ processes}\label{DF1}

It is instructive to
divide the $\Delta F = 1$ FCNC processes into
two classes. The first proceeds mainly through
QCD penguin diagrams {\it e.g.} $B\to K^0 \bar K^0, \phi K_S$ etc. The second is mediated
through electroweak penguin such as semi-leptonic
(i.e. $X l^+ l^-$)  decays and $\varepsilon'/\varepsilon\,.$
We show below that NP contributions
to processes from the first class are subleading. 
NP contribution to the second class of
processes are however comparable with the SM
ones.

Let us estimate the ratio between the NP and the SM contributions
for processes of the first class.
The leading NP contributions come from two sources.
The first is due to 
exchange of KK gluons discussed above in the context of  
$\Delta F = 2$ processes.
As an example we compare the $b\to s\bar ss$ NP contribution to
the SM one.
In term of spurions the leading NP contribution to 
this process is proportional to~(\ref{Lg}) 
\beq
{A_{b\to s\bar ss}^{\rm RS}}&\propto& 
\left( F_Q F^{\dagger}_Q \right)_{23}\approx
\sum_i\left(D_{L}^\dagger\right)_{2i} f_{Q^i}^{-2} \left(D_{L}\right)_{i3}  \nonumber\\
&\sim& C_s \left|V_{tb}^* V_{ts}\right| f_{Q^3}^{-2} 
\label{DelF1QCD}
\eeq
where $C_s$ is an order one complex number and as in the above
contribution from the RH sector, proportional to $f^{-2}_{d^i}$, 
are further suppressed.
The ratio between the NP and the SM contribution is roughly given by
\beq
{{A_{s\bar s s}^{\rm RS}}\over A_{s\bar s s}^{\rm SM}}
     \sim
     {16\pi^2\over E_0(m_t)}\, {2\over g_2^2 }\, {M_W^2\over m_{\rm KK}^2} 
      \, f_{Q^3}^{-2}\sim{ 0.1}\left({3 {\rm TeV}\over m_{\rm
           KK}}\right)^2
     \left({3\over { f_{Q^3}}}\right)^2 \,,\label{DeltaF1QCD}
\eeq

where $E_0(m_t)\sim0.3$ comes from computing the penguin diagram~\cite{Bur}.
The reason for the smallness of the NP contribution stems
from the fact that the flavor diagonal piece in the KK gluon coupling
yields a suppression of $1/\sqrt{k\pi r_c}\sim 0.15$~[see (\ref{Lg})].
%
\footnote{Note that, as mentioned earlier, reference \cite{Burdman:2003nt}
assumed smaller values of $f_{ Q^3 }$ such that the constraint from
$Z \rightarrow b \bar{b}$ is not satisfied. So, NP contribution
to $\Delta F = 1$ from KK gluon exchange can be larger than in our analysis.}
There is an additional contribution (which is further suppressed)
from the shift in the $Z$ couplings~(\ref{LZ}) which yields
${{A_{s\bar s s}^{\rm RS}}\over A_{s\bar s s}^{\rm SM}}\sim {g_2^2\over g_s^2}\sim0.2\,.$

Furthermore, one should have in mind the fact
that the above calculation was done at the EWSB scale.
When applying the evolution from the weak scale down to the the $m_b$
scale one finds that most of the contribution to the corresponding
Wilson coefficients is actually due to mixing with the tree level
operators~\cite{Bur}.
Therefore, to have a sizable NP effect one actually requires
the ratio in~(\ref{DeltaF1QCD}) to be of ${\cal O}(10)$~(see {\it
  e.g.}~\cite{Dilut}).
The conclusion from the above discussion is that 
the NP contribution cannot compete with processes 
which, within the SM, are induced by
$\Delta F=1$ QCD penguin diagrams.

We shall now move to discuss a class of processes which in the SM arise
only in the presence of electroweak penguin and box diagrams.
Examples of such processes are semileptonic $B,K\to X_{s,d}l
\bar l\,,\,B,K\to l \bar l\,,\,\varepsilon' /\varepsilon$.
We expect that NP contributions from shift of the coupling to the $Z$ will be comparable
to the SM ones for processes in the above class.
We choose to focus on $B\to X_sl^+ l^-$ and $K \to\pi \nu\bar\nu$ 
and show that indeed they receive sizable
NP contribution which might be measured in the near future.

We start with $b\to sl^+l^-$, in this case following the
literature~\cite{Zbs},
where we parameterize the contributions in term of the effective
 $Z$ flavor violating couplings, $Z_{sb}^L\,$. 
In term of spurions the leading NP contribution
is proportional to~(\ref{LZ}) 
\beq
Z_{sb}^{L\;{\rm RS}}&\propto& 
\delta g_Z\left( F_Q F_Q^\dagger \right)_{23}\approx
\delta g_Z\sum_i\left(D_{L}^\dagger\right)_{2i} f_{Q^i}^{-2} \left(D_{L}\right)_{i3}  \nonumber\\
&\sim& \delta g_Z C_z \left|V_{tb}^* V_{ts}\right| f_{Q^3}^{-2} 
\label{DelZbs}
\eeq
where $C_z$ is an order one complex number and contributions to the coupling
of the other
chirality,
$Z_{sb}^{R\;{\rm RS}}$,
are subdominant.  
The ratio between the NP and the SM contribution is then roughly given by
\beq
{Z_{sb}^{L\;{\rm RS}}\over Z_{sb}^{L\;{\rm SM}}}
     \sim
     {4\pi^2\over g^2_2 C_{10}^*}\,   {M_Z^2\over m_{\rm KK}^2}\, 
       {k\pi r_c \over f_{Q^3}^{2}}\sim{ 0.4}\times\left({3 {\rm TeV}\over m_{\rm
           KK}}\right)^2
     \left({3\over { f_{Q^3}}}\right)^2 \,,\label{DeltaZbs}
     \eeq
     where $C_{10}^*\sim1$ is the corresponding Wilson
     coefficient~\cite{Zbs}.
     This implies that within our framework order one deviation for
     the branching ratio of $b\to s l^+ l^-$
     from the SM prediction is expected \cite{talk, nomura2}. 
Similar deviation is also
     expected in the short distance contributions for the
     corresponding exclusive modes.
     The current experimental and theoretical uncertainties for the
     above inclusive branching
     ratio is of ${\cal O}\left(30\%\right)$ and
     ${\cal O}\left(20\%\right)$ respectively~\cite{Zbs}. It is clear 
     that the experimental statistical error is going to decrease in
     the near future. The above process therefore will yield an important
test for the framework.

Additional piece of non-trivial information can be extracted
by measurement of the lepton forward-backward asymmetry and 
spectrum of leptons \cite{talk}.
As the new physics (see Eq. \ref{DeltaZbs})) contributes  
mostly to the axial part of the lepton pair current it is expected
to yield a sizable modification to the angular distribution of
the outgoing leptons accompanied by a vector strange meson.
This implies modification of the location of the zero in the low $q^2$
region and the value of the integrated asymmetry for the high $q^2$
range~\cite{Zbs}. 

%
Note that the NP contribution has, in general, a weak phase not related to
SM contribution and also NP contribution has no strong phase 
from real intermediate states (unlike the SM
penguin diagram)
since it is a tree-level effect. Thus, $O(1)$ direct CP asymmetry
is expected in this process.

Similar flavor violating $Z$ coupling contributes also to
the $K^+\to\pi^+\nu\bar\nu$ decay process where for
this case we focus on the $Zsd$ coupling,
$Z_{sd}^{L\;{\rm RS}}$. In term of spurions we find
\beq
Z_{sd}^{L\;{\rm RS}}&\propto&
\delta g_Z\left( F_Q F_Q^\dagger \right)_{12}\approx
\delta g_Z \sum_i\left(D_{L}^\dagger\right)_{1i} f_{Q^i}^{-2} \left(D_{L}\right)_{i2}  \nonumber\\
&\sim& \delta g_Z\left[\left(D_{L}^\dagger\right)_{13} f_{Q^3}^{-2}
  \left(D_{L}\right)_{23}+\left(D_{L}^\dagger\right)_{12} f_{Q^2}^{-2}
  \left(D_{L}\right)_{22}\right]\nonumber\\
  &\sim&  \delta g_Z C'_z \left|V_{td}^* V_{ts}\right| f_{Q^3}^{-2} 
\label{DelZsd}
\eeq
where $C'_z$ is an order one complex number.  
The ratio between the NP and the SM contribution is then roughly given by
\beq
{{A_{K\rightarrow \pi\nu\bar\nu}^{\rm RS}}\over {A_{K\rightarrow \pi\nu\bar\nu}^{\rm SM}}}
     \sim
     (4\pi)^2\,{ a_{\nu}\sqrt{a_{d_L}^2+v_{d_L}^2}\over g^2_2 X_0(m_t)} \,   {M_Z^2\over m_{\rm KK}^2}\, 
       {k\pi r_c \over f_{Q^3}^{2}}\sim{ 0.3}\times\left({2 {\rm TeV}\over m_{\rm
           KK}}\right)^2
     \left({3\over { f_{Q^3}}}\right)^2,\label{Kpnn}
     \eeq
     where $X_0(m_t)\sim1.6$ is the corresponding Wilson
     coefficient~\cite{Zbs}. Thus we find a sizable NP contribution
     uncorrelated
     with the SM one.
The theoretical accuracy for the SM prediction for the 
BR($K^+ \to \pi \nu \nu$) is around 5-15\% \cite{Buras98, Falk:2000nm}.
     The recent
     experimental result in~\cite{E949} already probes the interesting
     region in the model parameter space in which the SM and the
     NP constructively interfere (though the future status of the
     experiment is not clear. For recent work see~\cite{Burasnu} and
     refs. therein).
Indeed the central value of the experimental result~\cite{E949} is
considerably higher than the SM prediction~\cite{Buras98, ANS_CKM} which 
could be an indication of NP;
however the errors are quite large at this point to draw a firm
conclusion. It is clearly very important to improve this
experimental determination.  

Indeed NP due to WED has even more striking implication for 
$K_L \to \pi^0 \nu \nu$. First recall that this process is theoretically
extremely clean as the rate in the SM is CP violating
and this is a very nice way to measure $\eta$ which drives
the CP violation in the CKM-paradigm~\cite{ckm}.  
Therefore, it is overwhelmingly dominated by the top quark. 
%
%
The presence of
the complex coefficient in eq.(\ref{Kpnn}), and therefore a new CP
odd phase, can also contribute in our framework 
to $K_L \to \pi^0 \nu \nu$
and cause a significant deviation from the prediction of the SM. 
%
%
These very difficult $K_L$ experiments~\cite{kopio}
then become a very nice way to constrain the phase
due to NP. The possible presence of NP source of CP in K-decays
due to WED should give an additional impetus to a separate
determination of the unitarity triangle purely  
from K-physics~\cite{Soni:2003rf}.   

     Using eqs. (\ref{Lg},\ref{LY},\ref{LZ}) it is straight forward to
     apply the above analysis to other similar $\Delta F=1$ electroweak
     processes. We shall not elaborate on the others since we hope
     that the procedure is rather transparent.
For recent overview on the current experimental and theoretical status
of other related processes see {\it e.g.}~\cite{Buras04} and refs. therein.

\subsection{Dipole operators}\label{Dipole}

The class of FCNC processes related to radiative B decays
provides a lot of
information on the structure of the effective theory
at and above the EWSB scale. 
Measurements of the BR$(b\to s\gamma)$
already provide a powerful constraint on NP models.
SM contributions
induce, to leading order, single chirality operators
$O_{7\gamma}$
and  $O_{8g}$ with 
$$O_{7\gamma,8g}=\bar b_R\sigma^{\mu\nu}s F^{\mu\nu},G^{\mu\nu}\,,$$
where $F^{\mu\nu}$ and $G^{\mu\nu}$ are the field strengths for the
electromagnetic and chromomagnetic interactions.
In the SM, the Wilson coefficients $C'_{7\gamma,8g}$ of the opposite
chirality operators, $O'_{7\gamma,8g}$, are suppressed by $m_s/m_b$
and therefore are negligible (or $m_d/m_b$ for the $b\to d\gamma$ processes).
The fact that there are no right handed chiral operators is a unique
feature to the SM which will be tested in the near future as discussed
below.
We therefore focus on the NP contributions for the above processes.
We show that in our framework the value of $C'_{7\gamma,8g}$ 
is found to be comparable with that of the SM contribution to $C_{7\gamma,8g}\,.$

From eqs. (\ref{Lg},\ref{LY},\ref{LZ})
we find that there are no tree level flavor changing contributions
which induce helicity flip as required by the above process.
Consequently, we now discuss loop processes which involve KK
states in the loop.
Contribution from an
individual KK state is finite but the sum is logarithmically sensitive to the
cutoff. This can be understood from the fact that the contribution must
involve the TeV brane (in the model with EWSB localized on TeV brane) to account
for the helicity flipping. This softens the UV sensitivity:
brane-localized interactions have higher degree of divergence than bulk interactions.
Consequently in the 5D Lagrangian there is a counter term
which cancels the above log divergence. 
Generically, without fine
tuning,
the finite loop contribution is not expected to be exactly canceled
by the counter term\footnote{As expected, the NDA estimate
for contribution from cut-off physics (which can be at tree-level) is
comparable to
the calculable loop contribution: 
see section \ref{EDM}.}. Thus our calculation below just estimates the
size of the finite UV insensitive NP contribution.
We also mention that with the low cutoff at the TeV brane, the logarithm
is not expected to be large and therefore,
is, in principle, under control.
As expected, with Higgs in bulk (but with wavefunction peaked
near TeV brane) \cite{bulkhiggs}, 
the loop contribution is finite 
since the helicity flip is no longer a brane-localized
interaction
and the cut-off physics contribution is smaller
(see section \ref{EDM}).

We can divide the leading NP contribution into three classes:
\begin{itemize}
\item[(i)] Flavor violating contributions due to zero mode
  couplings to KK gluons (shown in figure 4). These contributions are found to be subleading.
  \item[(ii)] Flavor violating contributions due to zero mode
    couplings to down type fermion KK modes which requires EWSB (shown in figure 5). These contributions are
    sizable even in the limit $\lambda_{u\,5D}\to0$.
 \item[(iii)]  Flavor violating contributions due to zero mode
    couplings to down and up type fermion KK modes which requires EWSB
    (shown in figure 6). These contributions are
    sizable but require non-zero up and down type Yukawa couplings.
    \end{itemize}
    Naively, it is not expected that loop processes with internal
    heavy KK fields would yield sizable contributions.
    The reason that we do find sizable contributions for the right handed
    operators
    stems from the fact that the entries of $D_R$ are all of order
    unity~(see table \ref{fstab}) \footnote{and 
couplings of KK fermions
to Higgs are enhanced 
by large $5D$ Yukawa required, in turn, 
to obtain top quark mass.} which overcomes the corresponding
    CKM suppression in the SM contribution.
    Below, the contributions (i,ii,iii) are discussed 
in that order.
    

    We start by
    showing why the contribution with KK gluons in the loop to
    $C'_{7\gamma}$, ${C'}^{G^{\rm KK}}_{7\gamma}$ are small.
    The relevant diagram is shown in fig. 4.
    In term of spurions it is proportional to
    \beq
    {C'}^{G^{\rm KK}}_{7\gamma}\propto \left( F_Q \lambda_{d\, 5D}
    F_d \right)_{32}
  \approx {1\over2v}\left[{\rm diag} \left(m_{d,s,b}\right) \right]_{32}=0\,,\label{bsgKKG}
  \eeq 
  where in the above we suppressed the flavor indices.
The reason why ${C'}^{G^{\rm KK}}_{7\gamma}$ vanishes is because,
    according to our
    approximation, it is aligned with the 4D
    down Yukawa matrix. This is discussed in detail in 
subsection \ref{relation} [see (\ref{align})],
whereas in appendix ~\ref{appkkfermion} we consider the
  deviations from the above result and show that the 
  corrections are indeed small.

\fmfcmd{%
vardef cross_bar (expr p,len,ang)=
((-len/2,0)--(len/2,0))
rotated (ang+angle direction length(p)/2 of p)
shifted point length(p)/2 of p
enddef;
style_def crossed expr p = 
cdraw p;
ccutdraw cross_bar (p, 5mm, 45);
ccutdraw cross_bar (p, 5mm, -45)
enddef;}

\begin{center}
\begin{eqnarray}
\parbox{70mm}{
\begin{fmfgraph*}(200,250)
\fmfleft{p1}
\fmfright{p2}
\fmftop{p3}
\fmf{fermion,tension=3.1,label=$s_R$}{v9,p2}
\fmf{plain,tension=5.5,label=$F_d$}{v2,v9}
\fmf{fermion,tension=3.1,label=$b_L$}{p1,v10}
\fmf{plain,tension=5.5,label=$F_Q$}{v10,v1}
\fmf{photon,tension1.5,label=$\gamma$}{p3,v3}
\fmf{gluon,tension=.9,label=$G^{(n)}$}{v1,v2}
\fmf{fermion,tension=.8}{v1,v45}
\fmf{plain,tension=5,label=$d^{(n)}_R$}{v45,v4}
\fmf{crossed,tension=1,label=$\lambda_{d~5D}^\dagger$}{v4,v5}
\fmf{fermion,tension=.7,label=$d^{(l)}_L$}{v5,v3}
\fmf{plain,tension=.7}{v3,v6}
\fmf{fermion,tension=1,label=$d^{(m)}_L$}{v6,v7}
\fmf{plain,tension=.8}{v7,v75}
\fmf{plain,tension=5}{v75,v2}
\end{fmfgraph*} } \nonumber
\end{eqnarray}
\vspace*{-4cm}\\
{\small Fig. 4: Contributions to $b\to s\gamma$ from KK gluon in the loop.}
\end{center}

  We next move to discuss the contributions with KK fermions in the
  loop (ii)+(iii).
  In term of spurions they are given by
  \begin{eqnarray}
  C'_{7\gamma}&\propto&\left[ F_Q  \left(
 \lambda_{u\, 5D }\lambda_{u\, 5D }^\dagger
+
\lambda_{d\, 5D }\lambda_{d\, 5D }^\dagger\right) \lambda_{d\, 5D }
 F_d \right]_{32}\label{spurGen}
\,,\end{eqnarray}
where the first (second) term in the parenthesis comes from computing the
diagram in fig. 6 (fig. 5).

Before  continuing we note that in the Higgsless case there are
 similar contributions as follows: the Higgs
 line should be replaced by a longitudinal $W,Z$ line which is
$W_5$, $Z_5$ in the Higgsless models.
In addition, just as in the model with the Higgs, the
 presence of the mass term on the TeV brane yield mixing between SM
 doublet (singlet) quarks and singlet (doublet) KK quarks.
Using this mass mixing, we can show that 
similar contributions arise also in the Higgsless case.

  In appendix \ref{ADipole} we show that these are not aligned with 
the 4D Yukawa matrices. Consequently
  we expect them to yield sizable contributions.
  In appendices \ref{DipoleGeneric} and \ref{apphiggsbsg},
  we computed the flavor structure 
  yielded by the diagrams shown in figures 5 and 6 respectively.
  It turns our that both the contributions from only down type 
(see Eq. (\ref{bsgdown}))
  and the up type (see
Eq. (\ref{bsgup})) internal KK quarks are comparable.
  We find that both of them are proportional to $m_b \left(D_R\right)_{32}$
 \beq
\hspace*{-.cm}   C'_{7\gamma}&\propto&a_\gamma
 \left(  \lambda_{ 5D } k \right)^2 { {m_{b}}} 
\left(D_R\right)_{23} 
\,,
  \eeq
  where $a_{\gamma}$ is an  order one complex number.
 In order to estimate the size of the contribution we divide it by the
 SM contribution, $C_{7\gamma}^{\rm SM}$
 \beq
{C'_{7\gamma}\over C_{7\gamma}^{\rm SM}}\sim {M_W^2\over m_{\rm KK}^2}\,
\left|{\left(D_R\right)_{23} \over V_{tb} V_{ts}^*}\right|\,
{\left( \lambda_{ 5D } k \right)^2 \over D'_0(m_t) g_2^2}\sim{ 1}\times\left({3 {\rm TeV}\over m_{\rm
           KK}}\right)^2\,
     {\left(D_R\right)_{23}\over 2}\,,\label{bsg}
 \eeq
 where $D'_0(m_t)\sim0.4$~\cite{Bur} and in the above estimate
 we used $\left(D_R\right)_{23}\sim f_{d^3}/f_{d^2}\sim
 0.5$ (see ~(\ref{fs})) and $\lambda_{ 5D } k \sim 4$ which
was required to obtain the top quark mass given the lower bounds
on $f_{ u^3 }$ and $f_{ Q^3 }$ as explained earlier.
The NP contributions to $C_{7\gamma,8g}$ 
are obtained by replacing $\left(D_R\right)_{23}$ in the above by
$\left(D_L\right)_{23}\sim V_{ts}$ and hence
are suppressed by $\sim {\cal O} \left( 1/10 \right)$.

 The above results imply that within our framework we expect the
  emitted photon
  to be affected by operators with both chiralities unlike the SM
  case. 
Upcoming results from the B factories will be sensitive to that
effect by either measuring the polarization of the outgoing
  photon~\cite{gross}
or by measuring time-dependent CPV effect ~\cite{ags}
in exclusive final states
such as $B_d \to \gamma K_s^{*0} (\rho, \omega)$ or in
$B_s \to \gamma \phi (K_s^{*0})$.  Indeed, experimental feasibility
of this interesting test has recently been demonstrated~\cite{babar_ags}
and improved tests with greater luminosities at B and Super-B factories
are eagerly awaited.

  Using similar derivation we find the prediction for the opposite
  chirality Wilson coefficient, $C'_{7\gamma\;d}$, for the $b\to
  d\gamma$ process~(\ref{bdgam})
  \beq
{C'_{7\gamma\;d}\over C_{7\gamma\;d}^{\rm SM}}\sim {M_W^2\over m_{\rm KK}^2}\,
\left|{\left(D_R\right)_{13} \over V_{tb} V_{td}^*}\right|\,
{\left( \lambda_{ 5D } k \right)^2 \over D'_0(m_t) g_2^2}\sim{ 3}\times\left({2 {\rm TeV}\over m_{\rm
           KK}}\right)^2\,
     {\left(D_R\right)_{13}\over 6}\,.\label{bsgd}
     \eeq
     In this case the effect is even more dramatic since within the
  SM the other chirality operator is 
suppressed by ${\cal O}\left({m_d\over m_b}\right)\,.$

It is also useful to note that RS1 contribution to 
$C_{ 7 \gamma}$, which is
$\sim {\cal O} \left( 1/10 \right)$ of SM contribution,
has, in general, a different weak phase than SM contribution
and also does not have a strong phase from real
intermediate states (unlike the SM contribution).
Hence, the NP 
can also affect quite appreciably 
SM predictions, which are fairly
precise, for direct CP asymmetries~\cite{soares}
in inclusive as well as exclusive
final states. 
In particular recall that SM predicts very small
($\approx 0.6\%$) asymmetry in $b \to s \gamma$ transitions which is
about an order of magnitude below the current experimental bound
\cite{nakao}. Continued experimental effort at measurment of direct
CP asymmetries in all of the radiative modes is clearly very
important.     

Note that the NP contribution to $C^{ \prime }_{ 7 \; \gamma }$ does not
interfere with the SM contribution to $C_{ 7 \gamma }$ in the rate for
$b \rightarrow s \gamma$ so that $C^{ \prime \; RS }_{ 7 \; \gamma }
\sim 1/2 C^{ SM }_{ 7 \gamma }$ is sufficient to be consistent with the
measured BR since the theoretical/experimental uncertainties are
at the level of $\sim {\cal O} ( 10 \% )$ each~\cite{rev_gamma,nakao}.

\begin{center}
\begin{eqnarray}
\parbox{80mm}{
\begin{fmfgraph*}(270,270)
\fmfleft{p1}
\fmfright{p2}
\fmftop{p3}
\fmf{fermion,tension=1.7,label=$s_R$}{v9,p2}
\fmf{plain,tension=4.,label=$\lambda_{d~5D}F_d$}{v2,v9}
\fmf{fermion,tension=1.7,label=$b_L$}{p1,v10}
\fmf{plain,tension=4.,label=$F_Q\lambda_{d~5D}$}{v10,v1}
\fmf{photon,tension1.5,label=$\gamma$}{p3,v3}
\fmf{dashes,tension=.8,label=${H}$}{v1,v2}
\fmf{fermion,tension=.8}{v1,v45}
\fmf{plain,tension=5,label=$d^{(n)}_R$}{v45,v4}
\fmf{crossed,tension=1,label=$\lambda_{d~5D}^\dagger$}{v4,v5}
\fmf{fermion,tension=.7,label=$d^{(l)}_L$}{v5,v3}
\fmf{plain,tension=.7}{v3,v6}
\fmf{fermion,tension=1,label=$d^{(m)}_L$}{v6,v7}
\fmf{plain,tension=.8}{v7,v75}
\fmf{plain,tension=5}{v75,v2}
\end{fmfgraph*} } \nonumber
\end{eqnarray}
\end{center}
\vspace*{-4.5cm}
{\small Fig. 5: Contributions to $b\to s\gamma$ from 
Yukawa interactions with down type KK quarks.}\\

\begin{center}
\begin{eqnarray}
\parbox{80mm}{
\begin{fmfgraph*}(270,270)
\fmfleft{p1}
\fmfright{p2}
\fmftop{p3}
\fmf{fermion,tension=1.7,label=$s_R$}{v9,p2}
\fmf{plain,tension=4.,label=$\lambda_{d~5D}F_d$}{v2,v9}
\fmf{fermion,tension=1.7,label=$b_L$}{p1,v10}
\fmf{plain,tension=4.,label=$F_Q\lambda_{u~5D}$}{v10,v1}
\fmf{photon,tension1.5,label=$\gamma$}{p3,v3}
\fmf{dashes,tension=.8,label=${H^\pm}$}{v1,v2}
\fmf{fermion,tension=.8}{v1,v45}
\fmf{plain,tension=5,label=$u^{(n)}_R$}{v45,v4}
\fmf{crossed,tension=1,label=$\lambda_{u~5D}^\dagger$}{v4,v5}
\fmf{fermion,tension=.7,label=$u^{(l)}_L$}{v5,v3}
\fmf{plain,tension=.7}{v3,v6}
\fmf{fermion,tension=1,label=$u^{(m)}_L$}{v6,v7}
\fmf{plain,tension=.8}{v7,v75}
\fmf{plain,tension=5}{v75,v2}
\end{fmfgraph*} } \nonumber
\end{eqnarray}
\end{center}
\vspace*{-4.5cm}
{\small Fig. 6: Contributions to $b\to s\gamma$ from 
Yukawa interactions with down and up type KK quarks.}

\subsection{Flavor diagonal CPV and electric dipole moments}\label{EDM}
It is well known that almost any SM extention contains new sources of
CPV.
Such sources generically contribute to two classes of CPV observables.
The first is related to what we discussed above, that is CPV that occurs
through
flavor mixing. Several experiments have measured CPV of the above
type in the K and B systems and it was found to be sizeable.
The second class is related to flavor diagonal CPV.
This type of CPV has not been observed yet and
experimental data yield a severe constraint on
the corresponding flavor diagonal CPV sources. This is through
measurements
of the electron and neutron electric dipole moments (EDMs), $d_n$.
Current data yield the following bound on the
neutron EDM~\cite{Harris}
\beq
d_n\leq6\times10^{-26}{\,\rm e\, cm}\,.\label{EDMexp}
\eeq 
In the SM 
there are two such sources. First,
there is the celebrated strong CP phase,
$\bar \theta$, where the above constraint 
yields
$\bar\theta\lsim 10^{-10}$~(see {\it e.g.}~\cite{Nir,Peccei} and refs. 
therein).
In addition, the CP-odd phase in the CKM matrix~\cite{ckm}
can yield a non-vanishing value for the EDM beyond two
EW loops~\cite{shabalin,banks}. However,
this is estimated to be extremely small,
$\lsim {\cal O} \left( 10^{-30} e-cm
\right)$.
In SM extensions other sources usually exist which contribute 
to the EDMs and therefore must be small.
A well known example is the supersymmetric CP problem where
generically the MSSM predicts the EDMs to be two order of magnitude
larger than the current experimental limit~(see {\it e.g.}~\cite{Nir,Barger} and refs. therein).

We find it therefore very important to investigate what is
the RS framework prediction regarding the EDMs.
Since the structure of the lepton sector is model dependent
we chose to focus on the quark sector.
The relevant operator, $O_n$, which contributes
to the EDM is given  by
\beq
O_{d_n}=\bar d_L\sigma^{\mu\nu} d_R F_{\mu\nu}+h.c.\,,
\eeq
where EDMs are proportional to the imaginary part of
the corresponding Wilson coefficient, $C_{d_n}$.
As explained in section~\ref{Dipole} since this process
requires
helicity flip there is no tree level contributions from KK modes.
We shall, therefore, focus on the leading, one-loop, contributions.
The required analysis is, actually, similar to the one done above in
the context of radiative $B$ decays.
In order to estimate $C_{d_n}$ we should calculate the contributions
from the diagrams shown in figs. 4-6 after we change the external quarks 
 into $d_{L,R}\,.$

We start by analysing the contibution with internal KK gluons, 
${C}^{G^{\rm KK}}_{d_n}$, from the
diagram in fig. 4.
In term of spurions it is given by (\ref{bsgKKG})
 \beq
    {C}^{G^{\rm KK}}_{d_n}\propto kv\left(D_L^\dagger F_Q \lambda_{d\, 5D}
    F_d D_R\right)_{11}
  \approx {1\over2}\left[{\rm diag} \left(m_{d,s,b}\right) \right]_{23}=0\,,
\label{dnKKG}
  \eeq 
The fact is that the imaginary part of ${C}^{G^{\rm KK}}_{d_n}$ vanishes 
because it is approximately aligned with the 4D
    down Yukwa matrix [see subsection \ref{relation}].
In appendix~\ref{appkkfermion} and \ref{Ddn} we consider the
  deviations from the above result and show that these are subleading
 compared to the larger predictions which we discuss below.

We therefore focus on the contributions given by the diagrams in figures 5 
and
6 (again changing the external quarks into $d_{L,R}$).
In term of spurions these are of the form
\begin{eqnarray}
  C_{d_n}&\propto& 2 k^3 v \left[F_Q  \left(
 \lambda_{u\, 5D }\lambda_{u\, 5D }^\dagger
+
\lambda_{d\, 5D }\lambda_{d\, 5D }^\dagger\right) \lambda_{d\, 5D }
 F_d \right]_{11}\label{dn1}
\,.\end{eqnarray}
In appendix~\ref{Ddn} 
we show that $C_{d_n}$, generically, contains an unsuppressed imaginary
part which cannot be removed by a phase redefinition,
\begin{equation}{\rm Im}\left(C_{d_n}\right)\sim \left|C_{d_n} \right|\,.
\label{ImEDM}
\end{equation}
Given the above let us estimate what is the model prediction for
the EDM 
 \begin{eqnarray}\hspace*{-.5cm}
\left( {d_n\over e} \right)_{ KK } &\sim& {1\over6}\,{m_d\over16\pi^2}\,
\frac{ \left( 2 k\lambda_{5D} \right)^2 }{ m_{ KK }^2 } 
\sim  10^{-24}{\rm cm}
\left({2 k\lambda_{5D}\over 4}\right)^2 
\left({3{\,\rm TeV}\over m_{\rm KK}}\right)^2
\label{dnAbs}
\,.\end{eqnarray}
The above result is larger than the experimental bound (\ref{EDMexp}) by a factor of
${\cal O}$(20).
This implies that, with $m_{\rm KK}\lsim 3\,$TeV, our
framework is
confronted by a CP problem similar to the SUSY CP problem.
Several points are in order regarding the above result
(see appendix \ref{Ddn} for details):
\begin{itemize}
  \item The imaginary part comes from ``Majorana"-like phases and
    therefore appears already at the two generation level.
    \item However, 
with one generation, Majorana phases are absent/not physical so that
the contribution does require presence of mixing in
      both $D_L$ and $D_R$ and non-degeneracy of both $f_{Q, d}$'s. 
     This implies the existence of a CP violating rephase
     invariant quantity, the
     analog of the Jarskog determinant of the SM, but originating
from Majorana phases, i.e.,
not
from CKM-like phases unlike in the SM. A similar object is
     discussed in the context of SUSY in the lepton sector ~\cite{Jars}. In
     that case 
     a new rephase invariant quantity is found which
     requires both mixing and non-degeneracy for 2 generation of both
     left handed and right handed sleptons in order
to generate EDM.
     \item The CKM-like phases cannot contribute to the EDM. This is
       since the one loop contribution does not involve all the three
mixing angles in left or right-handed down sector
simultaneously so that we do not have sensitivity to
SM-like Jarlskog invariants \cite{Jars}.
       \item The above contributions are sensitive both to the overall
         size of the Yukawa couplings and the KK masses. This imply that they
         decrease faster with larger KK masses than other
         signals described above. We
         elaborate more on this point in the conclusions.
  \end{itemize}

In addition there is a contribution from higher-dimensional operator
on the TeV brane. Since this contribution is UV sensitive, we can 
only estimate it using naive dimensional analysis and compare it with
our result in Eq. (\ref{dnAbs}) as follows. We should
replace the suppression from the heavy KK mass, $m_{ KK }$,  by warped-down cut-off
(since operator is on TeV brane), $\Lambda
\sim \Lambda_{ 5D } e^{ - k \pi r_c }$, where $\Lambda_{ 5D }$ is the
(Planckian) $5D$ cut-off. Furthermore the loop factor $
\left( 2 \lambda_{ 5D } k \right)^2 / \left( 16 \pi ^2 \right)$
is expected to be replaced 
by an $O(1)$ number. This is since cut-off contribution can be
tree-level (unlike ones which come from KK modes).
Consequently we find
\begin{eqnarray}
\hspace*{-.4cm}\left( \frac{ d_n }{e} \right)_{ \Lambda } & \sim & C_{ \Lambda } \frac{ m_d }
{ \Lambda ^2 } 
  \sim  C_{ \Lambda } 10^{-24}{\rm cm}
\left({2 k\lambda_{5D}\over 4}\right)^2 
\left( { 10{\,\rm TeV} \over \Lambda } \right)^2\,,
\label{dnAbscutoff}
\end{eqnarray} 
where $C_{ \Lambda }$ is an arbitrary complex coefficient.

The cut-off scale is model-dependent:
\begin{eqnarray}
\Lambda_{ brane } & \sim & \frac{ 4 \pi } { 2 \lambda_{ 5D } k } m_{ KK } \nonumber \\
 & \sim & 10 \; \hbox{TeV} 
\left( \frac{4}{2 \lambda_{ 5D } k } \right) 
\left( \frac{ m_{ KK } }{ 3 \; \hbox{TeV} } \right)
\nonumber \\
\Lambda_{ bulk } & \sim & 
\left( \frac{ 4 \pi } { 2 \lambda_{ 5D } k } \right)^2 m_{ KK } \nonumber \\  
 & \sim & 30 \; \hbox{TeV} 
\left( \frac{4}{2 \lambda_{ 5D } k } \right)^2 
\left( \frac{ m_{ KK } }{ 3 \; \hbox{TeV} } \right),
\end{eqnarray}
where the subscripts ``brane'' and ``bulk'' on $\Lambda$ denote the models
with Higgs on TeV brane \cite{cust1} and in the bulk \cite{bulkhiggs}, respectively 
Hence, EDMs from cut-off physics 
are comparable to the naive loop contribution
of Eq. (\ref{dnAbs}) so that they exceed experimental 
limit by ${\cal O} (20)$ 
for Higgs on TeV brane. 
On the other hand allowing a rather simple modification of our
framework, with Higgs in the bulk (but localized near the TeV brane)~\cite{bulkhiggs},
the cut-off effects are comparable to experimental limit
and significantly smaller than the contribution induced by figure 5
and 6 given in eq. (\ref{dnAbs}).

\subsection{Flavor violation in up quark sector}
\label{up}
%
%
%
In order to estimate flavor violation in the up type
sector we need to consider $U_{L,R}\,.$
Using table \ref{fstab} and eq.~(\ref{fs}), we find that
$\left( U_R \right)_{ 12 } \sim
f_{ u^2 }  / f_{ u^1 } \sim O \left( 10^{-2} \right)$ 
is much smaller than
$\left( U_L \right)_{ 12 }\,.$
Furthermore,
$\left( U_R \right)_{ 23 } \sim f_{ u^3 } / f_{ u^2 } \sim 0.1$ is somewhat larger than
$\left( U_L \right)_{ 23 }$ and $\left( U_R \right)_{ 13 } \sim 10^{
  -3 } $ is also smaller than $\left( U_L \right)_{ 13 }$.

In analogy to $\Delta F= 2$ transition in down quark sector, 
KK gluon exchange 
gives the following contribution to $D^0-\bar D^0 $ mixing
[up to $O(1)$ complex coefficients]:
\begin{eqnarray}
M_{ 12 \; LL }^{ RS } & \propto & \Big[ 
\left( U_L \right)_{ 13 } f^{-2}_{ Q^3 } \left( U_L \right)_{ 23 }
\Big]^2\,, \nonumber \\
M_{ 12 \; RR }^{ RS } & \propto & 
\Big[ 
\left( U_R \right)_{ 13 } f^{-2}_{ u^3 } \left( U_R \right)_{ 23 }
\Big]^2\,, \nonumber \\
M_{ 12 \; LR }^{ RS } & \propto & 
 \Big[ 
\left( U_L \right)_{ 13 } f^{-2}_{ Q^3 } \left( U_L \right)_{ 23 }
\Big] \Big[ 
\left( U_R \right)_{ 13 } f^{-2}_{ u^3 } \left( U_R \right)_{ 23 }
\Big],
\label{deltac2}
\end{eqnarray}
where $LL$ denotes contribution to
$\left( \bar{u}_L \gamma^{ \mu } c_L \right)^2$ operator and so on.
The fact that $f_{ u^3 }^{-1}$ is sizeable (in addition
to $f_{ Q^3 }^{-1}$) results in violation of approximate flavor symmetries/RS-GIM
and hence the $LR$ and $RR$ operators are also enhanced (unlike in the down quark
case).
We compare the NP contribution
(dominated by $RR$ operator) 
to the short distance contribution in SM \cite{Burdman:2003rs}
\begin{eqnarray}
\frac{ M^{ RS }_{ 12 \; RR } }
{ M^{ SM }_{ 12 } } & \sim & \frac{ 32 \pi^2 }{ N_c } \frac{ M_W^2 }{ m^2_{ KK } } 
\frac{ m_c^2 M_W^2 }{ \left( m_s^2 - m_d^2 \right)^2 } 
\frac{ k \pi r_c }{ f^4_{ u^3 } } \frac{ g_s^2 }{ g^4 } 
\Big[
\frac{ \left( U_R \right)_{ 13 } \left( U_R \right)_{ 23 } }
{ V^{ \ast }_{ cs } V_{ cd } } \Big]^2 \nonumber \\
 & \sim & 100 \left( \frac{ 3 \; \hbox{TeV} }{ m_{ KK } } \right)^2 
\left( \frac{ 1.2 }{ f_{ u^3 } }\right)^4,
\end{eqnarray}
where we used $m_c \approx 1.2$ GeV and $m_s \approx 100$ MeV.
%
%
We see that
the NP contribution is larger than the 
SM short distance effect by $O(100)$.
However, the long distance SM contribution
can be larger than the short distance SM
contribution by $O(100-1000)$~\cite{md_long}. Also, current
experimental
limit~\cite{pdb} is still weaker 
than the long distance SM prediction. 
So, at present there is
no constraint on this 
NP effect
using the $D^0-\bar D^0$ mass difference, $\Delta m_D$.

On the other hand, 
the presence of the complex coefficient in eq.~\ref{deltac2} means that
WED endows the $D^0-\bar D^0$ oscillation a non-standard CP-odd
phase. The presence of such a phase should cause time-dependent
CP asymmetries, perhaps $O(10\%)$,
in $D^0$ decays which may be cleanly measured
via  decays to CP-eigenstates such as $D^0 \to K_S 
\pi^0 (\eta, \eta', \rho...)$ or $\phi \pi^0$, in complete analogy to
$\sin 2 \beta$ measurments via $B^0 \to \psi K_S$.  
Note that direct CP asymmetries in these modes potentially arise
through interference of penguin and tree graphs, a la~\cite{bss_79}
and are expected to be completely negligible since the penguin
contribution (inlcuding 
NP effect) is extremely suppressed. Thus the existing upper bounds of a
few percent  asymmetries in many such channels~\cite{pdb}
are easily understood. This discussion underscores the importance of
pursuing 
time dependent CP studies in charm factories, such as CLEO-c~\cite{cleo_c}. 
%

There are also RS1 contributions to flavor-changing
top quark decays, for example, $t \rightarrow c Z$ (analogous to flavor-violating
$Z$ vertex involving down quarks) and $t \rightarrow c \gamma$
(analogous to radiative $B$ decays).
A new effect in radiative decays is that, 
as mentioned earlier, $f_{u^3}$ for 
the top quark
is quite different than that of
all the other up type singlet quarks.
Consequently, the wavefunction and spectrum of $t_R$ KK modes is different than that of
other KK fermions and 
our approximation of KK-blind flavor violation breaks down.
This results in KK gluon contribution to dipole operators being 
{\em not} aligned
with $4D$ up Yukawa matrix and so it does contribute to $t \rightarrow c \gamma$
(see appendix). 
%
%
Recall that in the SM, due to the large mass of the top quark
the GIM-mechanism becomes exceedingly effective so that 
not only the branching ratios
of these decays ($t \to c Z(\gamma, gluon$)
are extremely suppressed~\cite{Eilam:1990zc,Grzadkowski:1990sm}, all CP
asymmetries~\cite{Eilam:1991yv, Atwood:2000tu}
driven by the CKM paradigm~\cite{ckm} 
become completely negligible.  
Since these decays are sensitive to
RS1 top quark FC coupling (including effect of new CP-odd phase(s)),
their searches are very well motivated.  
Such decays can be probed at the Tevatron, 
LHC and a linear collider (LC) -- we will
leave this analysis for a future study~\cite{APS_top_WIP} 
except also to draw brief attention to another unique process, 
$e^+ e^- \to \bar t c, t \bar c$. These are also very clean
reactions to study at a LC in search of signatures of RS1
as in the SM their rate is,
once again, exceedingly suppressed\cite{Atwood:1995ud}.  

\section{Discussion and conclusions}\label{Conc}

It is well known that in order to
solve the fine tuning problem any new physics (NP) framework is
required to introduce new degrees
of freedom at a scale close to the electroweak symmetry breaking
(EWSB) scale. These new weak-scale fields, however, tend to spoil 
the successful fit of the SM to
electroweak precision measurements (EWPM)  and to various measurements
related to flavor violation. Thus a tension is induced between
the SM experimental success and the need to solve the fine tuning problem.

Recently a framework (containing both Higgs and 
Higgsless models) based on warped geometry
with bulk custodial symmetry which relaxes the tension related to
EWPM was introduced.
We focused in this paper on the flavor sector of this framework
with light fermions localized near Planck brane
to make flavor issues UV-insensitive.
We showed that, regardless of the details of the model,
it has an underlying organized structure.
This framework
has a built-in mechanism to suppress NP contribution related to flavor
changing neutral currents (FCNC)s. 
It turns out that effectively flavor dynamics
is controlled only by physics near the TeV brane.
This implies that besides the SM Yukawa matrices 
all the NP flavor violation is governed by three additional
spurions $F_{Q,u,d}$ that transform as bi-fundamentals
under the SM and the diagonal KK quark flavor group,
U(3)$_{Q,u,d}\times~$U(3)$^{n}_{Q,u,d}\,.$
These are just related to the value of the zero mode profiles
on the TeV brane. The built in suppression of FCNC stems from the fact that:
(i) the entries in $F_{Q,u,d}$ related to the light quarks are small which
yield an approximate flavor symmetries. (ii) only the non-universal part in $F_{Q,u,d}$
can induce flavor violation which yield the RS-GIM mechanism. 
Consequently we find that, as in the SM,
flavor violation in the model is sizable due to third generation GIM breaking.

The actual models that we study provide a solution to the SM flavor puzzle.
That is the 5D Yukawa matrices are assumed to be anarchical and the
hierarchy in the SM flavor parameters is accounted by the split fermion
mechanism. This assumption turns the framework into a predictive one and
our results become robust and independent within a class of models in this
overall framework.   

We find that NP could be detected in three classes of FCNC process:
(i) $\Delta F=2$ transitions; (ii) $\Delta F=1$, mainly in
semi-leptonic decays (e.g. $B \to X_s l^+ l^-$); (iii) Radiative B decays.
In addition we showed the contributions to EDMs from KK states are 
about an order 
of magnitude above the current experimental bound for
KK masses $\sim 3$ TeV.
Thus there is an RS CP problem. 

It is important to note that the contributions related to class (iii) 
including the EDMs are more sensitive to the assumption that the
KK masses are small as follows. This is due to the fact
that they are 
proportional to 
both the square of the 5D Yukawa couplings
and inverse square of the KK masses.
If we slightly increase the KK masses (to
$\sim 4$ TeV) however the EWPM related to the
shift of the coupling of $b_L$
to $Z$ are weakened. This implies that $Q_3$ can localized
more towards the TeV brane ($f_{Q^3}$ is smaller)
which enhances the top mass.
Consequently this allows for a lower (by a factor
of $\sim 2$) overall scale
of the Yukawa couplings. 
Thus the new physics contributions to radiative B decays and EDMs
falls much more rapidly (by a factor of
$\sim 10$ for KK mass $\sim 4$ TeV) than naively expected.
On the other hand, 
the NP contributions of classes
(i+ii) 
are proportional to square of
$b_L$ coupling to {\em gauge} KK mode and to 
the inverse square of the KK masses: the latter increases
as $Q_3$ is localized closer to the TeV brane in such a way
that these NP effects remain roughly the same for KK mass $\sim 4$ TeV. 

We finally want to comment of how to proceed from this point.
Our work is based on the concept that we should test the predictions of
the framework
allowing for minimal possible fine tuning.
This is why we assume that the Yukawa matrices are anarchical
and that the KK masses are rather low to reduce the little hierarchy.

Future measurements of processes related to items (i)-(iii)
might find deviation from the SM prediction supporting the above
framework.
Two questions are important to ask in advance as follows:
\begin{itemize}
 \item What if no deviations from SM predictions are found? \\
We can either just say that the KK masses are actually
higher - this implies that the framework requires more fine tuning
and therefore becomes less attractive.
Other possibility is that it might be that accidentally one or two
elements of the 5D Yukawa matrices 
are smaller than their naive value. This may induce a small
value of $\left(D_L\right)_{13}$. The consequence of this is
accidental suppression of some of the above contributions.
We believe that due to correlation between various observables
discussed above one will be able to test this hypothesis and verify or
falsify this explanation. Analysis of non-trivial correlations between
the
predictions of 
this framework is left to future research.
\item What if one finds deviation in one of the classes of FCNC
  processes which are not sensitive to NP contribution within our
  framework?
  In this case we claim that the above framework will be disfavored
  and we do not find a way of accounting for such a situation.
  A realistic example of such a scenario is data which would signal
  a large deviation from the SM prediction in processes which in the
  SM are dominated by tree level or QCD penguin diagram. Leading candidate for
  that situation is the CP asymmetry in $B\to\phi K_S$.    
\end{itemize}
We also should remark that in our study above we consider the simplest
class of model without TeV brane kinetic terms~\cite{braneterm} and with
EWSB (with or without Higgs) on the TeV brane. 
Note, however, in 
models with Higgs
in the bulk
~\cite{bulkhiggs}, the Higgs profile is still localized near TeV brane.
So, our study is valid for these models as well.
We should though stress that
recent studies of electroweak precision tests (EWPT)
in warped Higgsless models
\cite{barbieri2, hewett2} suggest that, even in the presence of
brane kinetic terms, these models can be consistent with EWPT only
if the KK modes (even the 1st one) are strongly coupled.
This implies the presence of new physics at the
KK mass scale (related to the cut-off)
leads to
loss of predictivity since the NDA size of 
(uncalculable) cut-off effects are comparable to
(or in the case of radiative effects, larger than)
the KK effects which we studied.

\flushleft{\bf Acknowledgements}\\
KA~is supported by the Leon Madansky fellowship
and NSF Grant\\ P420D3620414350;
GP~is supported by the Director, Office of Science, Office of High
Energy and Nuclear Physics of the US Department of Energy under
contract DE-AC0376SF00098.
Research of AS is supported in part by
DOE under Contract No. 
DE-AC02-98CH10886. 
We thank organisers of 2nd Workshop on the 
Discovery Potential 
of a B Factory at $10^{ 36 }$ Luminosity,
SLAC (October, 2003) where
this work was initiated and
G.~Burdman, Z.~Chacko, W.~Goldberger, M.~Graesser, Y.~Grossman, I.~Hinchliffe, D.~E.~Kaplan,
Y.~Nir, Y.~Nomura, F.~Petriello, A.~Pierce, R.~Sundrum, M.~Suzuki and
J.~Wacker for discussions.

\section*{Appendix}
\begin{appendix}


\section{Coupling of zero-mode fermion to gauge KK mode}
\label{appzerogaugekk}

The couplings of KK and zero modes are given by overlap of their wavefunctions.
Decomposing the $5D$ gauge fields as
$A_{ \mu } 
( x, z ) = \sum_n A^{ (n) } (x) f_n (z)$, 
the wavefunction of gauge KK mode is given by
(for simplicity, we omit the index for the three SM gauge groups in the following)
\begin{equation}
f_n ( z ) = 
\sqrt{ \frac{1}{z_h} } \frac{z}{ N_n } 
\left[ J_1 \left( m_n z \right) + b_n
Y_1 \left( m_n z \right) \right] \,.
\label{fn++}
\end{equation}
We can show that it is
{\em localized near the TeV brane}.
We first consider gauge field with $(++)$ boundary condition  
(in the Higgs models all the SM gauge boson have these boundary conditions), 
for which
the normalization factor, $N_n$, is given by
\begin{eqnarray}
N_n^2 & = & \frac{1}{2} \left[ z_v^2 \big[ J_1 \left( m_n z_v \right) + b_n
Y_1 \left(  m_n z_v \right) \big]^2 - 
z_h^2 \big[ J_1 \left(  m_n z_h \right) + b_n
Y_1 \left(  m_n z_h \right) \big]^2 \right]   
\nonumber \,.
\end{eqnarray}
The masses of gauge KK modes and $b_n$ are found from
\begin{eqnarray}
\frac{ J_0 \left( m_n z_h \right) }{ Y_0 \left( m_n z_h \right) } & = & 
\frac{  J_0 \left( m_n z_v \right) }{  Y_0 \left( 
m_n z_v
\right) }= - b_n\,.
\end{eqnarray}
We will need masses of lightest KK modes only so that henceforth
we assume $m_n z_h \ll 1$. Then,  
we get
$m_n z_v \approx $ zeroes of $J_0 + O \left( 1 / 
\big[ 
\log m_n z_h 
\big] \right)$. In particular,
the mass of the lightest gauge KK mode is given by
\begin{equation}
m_1 \approx 2.45 z_v
\label{lightestgaugeKK}
\end{equation}
For $m_n z_v \approx $ zeroes of $J_0 \gg 1$, i.e., $m_n z_v
\approx \pi \left( n - 1/4 \right)$,
we find that
\begin{eqnarray}
N_n^2 & \approx  & \frac{ z_v }{ \pi m_n } 
\end{eqnarray}

As in the case of a {\em flat}
extra dimension, the zero-mode of gauge field 
(which is identified with the SM gauge boson)
has a flat profile
so that its couplings to all particles is given, at tree level, by
\begin{equation}
g = g_{ 5D } / \sqrt{ \pi r_c }
\label{0mode}
\end{equation}
where $g_{ 5D }$ is the $5D$ gauge coupling.
In terms of this $4D$/SM gauge coupling,
the coupling of zero-mode fermion to gauge KK mode (in
the interaction basis) is given by:
\begin{eqnarray}
\frac{ g^{ (n) } (c) }{ g  } & = &
\sqrt{ \pi r_c }
\int dz \sqrt{-G} \; \frac{z}{z_h} \; \chi_0^2 ( c, z ) f_n (z), 
\label{zero1kk}
\end{eqnarray}
where
$z / z_h$ is the funfbein factor and $-G = 
\left( z / z_h \right)^{-5}$ is the determinant of the metric
and
the wavefunction of the fermion zero-mode is given by
\begin{equation}
\chi_0 (c, z) = \sqrt{ 
\frac{1-2c}{ z_h \left( e^{ k \pi r_c ( 1-2c) }
-1 \right) } } \left( \frac{z}{z_h} \right)^{2-c}
\,.
\end{equation}
Recall
that the ``canonically'' normalized fermion zero-mode wavefucntion
is given by $z^{ -3/2} \chi_0 (c, z)$.
A numerical evaluation of this coupling shows that it has the
form of 1st term in
Eq. (\ref{Lg}). 
The $\sim $ sign in Eq. (\ref{Lg}) stands for an additional 
$O(1)$ $c$-dependent factor.
For a CFT interpretation of this form of the coupling, see section 
\ref{cft}.

For completeness, we
now consider the combination
of $U(1)_R$ and $U(1)_{ B - L }$ which is orthogonal to hypercharge and
is
denoted by
$Z^{ \prime }$. It has $(-+)$ boundary condition (i.e.,
no zero-mode) and the normalization
factor for KK modes
of $Z^{ \prime }$ is given by
\begin{eqnarray}
N_n^2 & = & \frac{1}{2} \left[ z_v^2 \big[ J_1 \left( m_n z_v \right) + b_n
Y_1 \left(  m_n z_v \right) \big]^2 - 
z_h^2 \big[ J_0 \left(  m_n z_h \right) + b_n
Y_0 \left(  m_n z_h \right) \big]^2 \right]  
\nonumber \\
\end{eqnarray}
and the masses of gauge KK modes and $b_n$ are given by
\begin{eqnarray}
{\hspace*{-.2cm}}\frac{ J_1 \left( m_n z_h \right) }{ Y_1 \left( m_n z_h \right) } & = & 
\frac{  J_0 \left( m_n z_v \right) }{  Y_0 \left( 
m_n z_v
\right) } = - b_n,
\end{eqnarray}
so that, for $m_n z_h \ll 1$, we get
$m_n z_v \approx $ zeroes of $J_0$.
For 
$m_n z_v \; ( \approx \; \hbox{zeroes of} \; J_0 ) \gg 1$, i.e., $m_n z_v \approx
\pi \left( n - 1/4 \right)$, we can show
that
%
%
$N_n^2 \approx z_v / \left( \pi m_n \right)$ as before.

The couplings of fermions to $Z^{ \prime }$ KK mode can be obtained 
in a similar fashion.

The coupling of gauge KK modes to Higgs
is obtained by evaluating the wavefunction on TeV brane (which is (approximately) the same
for both $(++)$ and $(-+)$ BC):
\begin{eqnarray}
\frac{ 
g^{ ( n ) } _{ Higgs } 
}
{ g }
& \approx &
(-1)^{ ( n - 1 ) } 
\sqrt{ 2 k \pi r_c }
\label{gaugeKKHiggs}
\end{eqnarray}
%


\section{KK fermion wavefunction and spectrum}
\label{appkkfermion}

Expanding the $5D$ fermion as $\Psi ( x , z ) = \sum_n \psi^{ (n) } (x)
\chi_n ( c , z )$, the
wavefunction of KK mode of SM fermion (i.e.,
(++) boundary condition) with mass $m_n$ is given by:
\begin{eqnarray}
\chi_n (c, z) = \left( \frac{z}{ z_h } \right)^{5/2} \frac{1}{ N_n \sqrt{ \pi r_c } }
\,\left[ J_{ \alpha } \left( m_n 
z \right) + 
b_{ \alpha } (m_n) 
Y_{ \alpha } \left( m_n z
\right)
\right],
\end{eqnarray}
where 
$\alpha = | c  + 1/2 |$ and
$m_n$ and $b_{ \alpha}$ are given by 
\begin{eqnarray}
\frac{ J_{ \alpha \mp 1} \left( m_n z_h \right) }
{ Y_{ \alpha \mp 1} \left( m_n z_h \right) } & = & 
\frac{ J_{ \alpha \mp 1} \left( m_n z_v \right)  }
{ Y_{ \alpha \mp 1} \left( m_n z_v \right) } \equiv - b_{ \alpha } (m_n), 
\label{tR++}
\end{eqnarray}
(with upper (lower) signs for $c > -1/2$ ($c < -1/2$))
and
\begin{eqnarray}
N_n^2 & = & \frac{1}{ 2 \pi r_c z_h } 
\left[ z_v^2 \big[ J_{ \alpha } \left( m_n z_v \right) 
+ b_{ \alpha } (m_n) Y_{ \alpha } \left( m_n z_v \right) \big]^2\right. \nonumber\\ 
&&-\left.z_h^2 \big[ J_{ \alpha } \left( m_n z_h \right) 
+ b_{ \alpha } (m_n) Y_{ \alpha } \left( m_n z_h \right) \big]^2 \right]\,.\label{KKF2}
\end{eqnarray}
Just as for gauge KK modes, we will 
assume below
$m_n z_h \ll 1$ since we are interested in only lightest KK modes.
For $-1/2 < c < 1/2 - \epsilon$ (where $\epsilon 
\stackrel{>}{\sim} 0.1$), we get $m_n z_v \approx$ zeroes of $J_{ -c + 1/2 }
\approx \pi \left( n - c / 2 \right)$, where the last formula
is valid for $m_n z_v \gg 1$. 
Similarly, 
for $c > 1/2 + \epsilon$, we get $m_n z_v \approx$ zeroes of $J_{ c - 1/2 }
\approx \pi \left( n + c / 2 - 1/2 \right)$, where the last formula
is valid for $m_n z_v \gg 1$. 

The wavefunction of KK mode
of $u^{ \prime }_R$ and $d^{ \prime }_R$ [$SU(2)_R$ partners of SM $u_R$ and $d_R$
with $(-+)$ boundary condition] with mass $m_n$ is similar
to those 
for $(+,+)$ KK modes, except:
\begin{eqnarray}
\frac{ J_{ \alpha } \left( m_n z_h \right) }
{ Y_{ \alpha } \left( m_n z_h \right) } & = & 
\frac{ J_{ \alpha \mp 1} \left( m_n z_v \right) }
{ Y_{ \alpha \mp 1} \left( m_n z_v \right) } = - b_{ \alpha } (m_n). 
\end{eqnarray}
and
\begin{eqnarray}
N^2_n & = & 
\frac{1}{ 2 \pi r_c z_h } 
\Big[ z_v^2 \big[ J_{ \alpha } \left( m_n z_v \right) 
+ b_{ \alpha } (m_n) Y_{ \alpha } \left( m_n z_v \right) \big]^2
\nonumber\\ && -
z_h^2 \big[ J_{ \alpha \mp 1 } \left( m_n z_h \right) 
+ b_{ \alpha } (m_n) Y_{ \alpha \mp 1 } \left( m_n z_h \right) \big]^2 \Big]
\,.\label{tR+-}
\end{eqnarray}
For $c > -1/2 + \epsilon$, 
$m_n z_v \approx$ zeroes of $J_{ c-1/2 } \approx \pi \left( n - 1/2 + c/2 \right)$, 
where the last formula
is valid for  $m_n z_v \gg 1$. 

We can show that, for both types of KK 
fermions and for $m_n z_v \gg 1$
\begin{eqnarray}
N_n^2 & \approx & 
\frac{ z_v }{ z_h }\, \frac{1}{ \pi^2 m_n r_c }\label{tR+-1}
\end{eqnarray}
and that the {\em wavefunctions are localized near the TeV brane} (just as
for gauge KK mode).

\subsection{Couplings of KK fermions to Higgs}
 
Using eqs. (\ref{tR++}-\ref{tR+-1}) one can show 
that the value of the KK fermion wavefunctions on the
TeV brane is roughly given by
$\sqrt{ 2 k }$
so that
\begin{eqnarray}
\left.\frac{ \hbox{wavefunction of KK fermion} }{ \hbox{wavefunction of 
(SM) zero-mode fermion} }
\right|_{ \rm{TeV\, brane} } & \approx & f \left( c \right)
\end{eqnarray}
Hence, in the interaction basis, 
we get
\begin{eqnarray}
Q_{ i }^{ (n) } d_{ j }^{ (m) } H
\; \hbox{coupling} & \approx & 2 \lambda_{ 5D
\; ij } k
\end{eqnarray}
and similarly for other KK modes.
Also, 
\begin{eqnarray}
Q_{ i }^{ (0) } d_{ j }^{ (n) } H
\; \hbox{coupling} & \approx & \frac{ 2 \lambda_{ 5D
\; ij } k }{ f_{ d^i } }
\label{zerokkhiggs}
\end{eqnarray}
and similarly other couplings.

These couplings 
appear in Eq. (\ref{LY}).
In particular, this explains $\approx$ sign in Eq. (\ref{LY}), i.e.,
there is
no further $O(1)$ flavor-dependence (cf. coupling to gauge KK mode 
in Eq. (\ref{Lg})).

\subsection{Couplings of KK fermions to gauge KK modes}

The coupling of zero-mode fermion in {\em weak} eigenstate
basis to {\em KK} fermion and gauge {\em KK} mode is diagonal in generation
space and is given by:
\begin{eqnarray}
\frac{ g^{ (0, n, m) } (c) }{ g } & = &
\sqrt{ \pi r_c }
\int dz 
\sqrt{-G} \; \frac{z}{z_h} \; \chi_0 ( c, z ) \chi_n ( c , z ) f_m (z), 
\label{zero2kk}
\end{eqnarray}
where, as before, 
$z / z_h$ is the funfbein factor and $-G = 
\left( z / z_h \right)^{-5}$ is the determinant of the metric

Similarly, we can obtain the coupling of $2$ KK fermions to gauge KK mode:
\begin{eqnarray}
\frac{ g^{ (m, n, p) } (c) }{ g } & = &
\sqrt{ \pi r_c }
\int dz 
\sqrt{-G} \; \frac{z}{z_h} \; \chi_m ( c, z ) \chi_n ( c , z ) f_p (z)
\end{eqnarray}
A numerical evaluation of these wavefunction overlaps shows that
these couplings have the form of 2nd and 3rd terms of Eq. (\ref{Lg}).
Just like for 1st term of
Eq. (\ref{Lg}) (i.e.,
coupling of $2$ zero-mode fermions to gauge KK mode), 
the $\sim $ sign in Eq. (\ref{Lg}) stands for an {\em additional} 
$O(1)$ $c$-dependent factor (cf. coupling to Higgs in Eq. (\ref{LY})).
For a CFT interpretation of the form of these couplings, see section
\ref{cft}.


\subsection{Flavor structure of KK gluon diagram}


Consider the KK gluon contributions to flavor-violating dipole operators, i.e.,
$C^{ \prime }_{ 7 \gamma }$ and $C_{ 7 \gamma}$:
\begin{eqnarray}
C^{ \prime }_{ 7 \gamma } & \propto & v\left[ D^{ \dagger }_L\, {\rm diag}\Big(a_{Q^i} f^{-1}_{Q^i} \Big)
\lambda_{ d \, 5D }\, 
{\rm diag}\Big( a_{d^i} f^{-1}_{d^i} \Big) D_R \right]_{ 32 } \nonumber \\
C_{ 7 \gamma } & \propto & v\left[ D^{ \dagger }_L\, {\rm diag}\Big( a_{Q^i} f^{-1}_{Q^i} \Big)
\lambda_{ d \, 5D } \,{\rm diag}
\Big( a_{d^i} f^{-1}_{d^i} \Big) D_R \right]_{ 23 } \,,
\end{eqnarray}
where the above expression is in the special interaction basis in which the
bulk masses are diagonal.
We have now included 
flavor dependence parameterized by the $O(1)$ coefficients, $a_{Q^i,d^i}$, (i.e., flavor-dependence
{\em apart} from $f$'s). These appear in the form of the exact couplings of
zero-mode fermion to KK fermion and KK gluon in Eq. (\ref{Lg}).

Numerical evaluations shows
that only the $a_{Q}$'s have $O(1)$ flavor dependence as follows.
It turns out that the KK wavefunctions for down-type quarks
are similar since $c_{Q, d}$'s 
are all close to 
$1/2$ (KK wavefunctions are not so sensitive to $c$'s).
However, since
$c_{ Q^3 } \sim 0.3-0.4$, the wavefunction of zero-mode $b_L$ is localized
a bit near TeV brane, whereas $c_{ Q^{ 1, 2 } } \sim 0.6-0.7$ so that wavefunction
of zero-modes are
localized near Planck brane.
This difference in {\em zero}-mode wavefunctions
results in $a_{ Q^{1,2 } } / a_{ Q^3 } \sim 1.5$.
Whereas {\em all}
$c_d
\approx 0.6-0.7$ so that all zero-modes
are localized near Planck brane resulting in very small
flavor-dependence in $a_d$'s.

Another effect that might spoil our approximation is splitting in
masses between the masses of the same level KK fermions.
Using the results of the previous section we find that this
splitting
is at most of ${\cal O} ( 10 \% )$ for both the doublet and singlet
down type quarks and therefore is subdominant for most
calculations.

Alltogether we find [up to  ${\cal O} ( 10 \% )$
corrections in 
$a_Q$
due to the mild splitting in KK mass] 
\beq
a_{Q^i} \sim O(1)\times \Big( 1.5\,, \, 1.5\,,\, 1\Big)\,,\qquad
a_{d^i} \sim O(1)\times \Big( 0.9 \,,\, 1\,,\, 1 \Big)\label{as}
\eeq
Note that $c_{d^2} \sim c_{ d^3}$ (see table \ref{fstab}) so that 
these two KK masses and hence the last two entries in $a_d$ are degenerate.

Thus, the approximation of neglecting flavor dependence in $O(1)$ coefficients
in 2nd term of Eq. (\ref{Lg}) is very good for the singlet down quarks and
is subject for ${\cal O}(1)$ corrections for the doublet ones.

Including
the above effect, we get
\begin{eqnarray}
C^{ \prime }_{ 7 \gamma } & \propto & 
m_s \left(V_{\rm CKM} \right)_{23} 
+ {\cal O} ( 0.1 )\times m_b \; \left( D_R \right)_{ 13 } 
\left( D_R \right)_{ 12 },
\end{eqnarray}
where 1st (2nd) contribution to 
$C^{ \prime }_{ 7 \gamma }$
is using left (right)-handed mixing.

Thus, NP contribution from KK gluon to $C^{ \prime }_{ 7 \gamma }$
is suppressed by roughly $\left( g_s \sqrt{ k \pi r_c } \right)^2 
M_W^2 / m_{ KK }^2
\sim {\cal O} \left( 1/10 \right)$
compared to SM contribution to same operator.
We conclude that the mild flavor dependence
in $a_d$'s 
does not give a significant effect in 
$b \rightarrow s \gamma$ due to near degeneracy of
$c_{ d^{2, 3}}$ (cf. $b \rightarrow d \gamma$ below).
Given eq. (\ref{as}) one find that the left handed chirality operator also receives
NP contributions
\begin{eqnarray}
C^{\rm RS}_{ 7 \gamma } & \propto & 
m_b \left(V_{\rm CKM }\right)_{ 23 }\,,
\end{eqnarray}
but without enhancment due to large right handed mixing
[compare with the leading contribution (\ref{bsg})]. Consequently we find that this 
NP contribution to $C_{ 7 \gamma }$ is suppressed by ${\cal O}(1/10)$ compared to
SM contributions.
We conclude that the
$O(1)$ flavor-dependence in $a_Q$ does not give interesting effect since mixing is 
similar to SM, i.e., CKM-like. 

Similarly, for $b \rightarrow d$ transition, we get
\begin{eqnarray}
C_{7 \gamma \; d }^{\rm RS} & \propto & m_b \left(V_{ CKM }\right)_{ 13 }
\nonumber \\
C^{ \prime }_{ 7 \gamma \; d } & \propto & \frac{1}{10} m_b \left( D_R \right)_{13}\,,
\end{eqnarray}
Consequently the
NP contributions to $C_{ 7 \gamma \; d}$ are of ${\cal O} ( 10 \% )$ 
of the size of SM contributions.

In the case of  $C^{ \prime }_{ 7 \gamma \; d }$ the situation is
different since the SM contribution are completely negligible. 
On top of that we know that right-handed mixing are
much larger than left-handed. Thus  we find that
the NP contributions to $C^{ \prime }_{ 7 \gamma \; d }$  
are enhanced even though the dispersion in
the $a_{d^i}$ is of ${\cal O} ( 10 \% )$. Plugging the actual values
we find the NP contribution are larger than the SM contribution
to the same operator by
${\cal O}(10) $. It is still only 10\%
of the leading SM contributions to $C_{ 7 \gamma d }$ 
(Of course, the leading contributions that we find in Eq. (\ref{bsgd}), 
are even larger, comparable to the SM leading ones).

Note that since the right-handed mixings are larger than left-handed,
the 
contribution 
to both $C^{ \prime }_{ 7 \gamma }$ and $C^{ \prime }_{ 7 \gamma \; d}$ 
would have been comparable to SM
contributions to $C_{ 7 \gamma }$
and $C^{ \prime }_{ 7 \gamma \; d }$,
respectively
if the $a_d$'s had  
${\cal O}(1)$ flavor dependence. This is exactly the case in the up
quark sector which we consider in the following.

There is a new effect in up-quark sector as follows. In addition to $O(1)$
flavor-dependence in $a_Q$ (same as down sector), there
is $O(1)$ flavor-dependence in $a_u$. This implies that
approximation of neglecting the flavor dependence in ${\cal O}(1)$ coefficients
in the 2nd term of Eq. (\ref{Lg}) is
not valid for both left and right-handed up quarks.
This is because
$c_{ u^3 }
\lsim 0$, i.e., much smaller than $c_{ u^{ 1, 2 } } $ 
which are $\approx 
0.6-0.7$. Thus, the zero-mode 
wavefunction of $t_R$ is localized near TeV brane, 
whereas $u_R$ and $c_R$ are localized 
near Planck brane. In addition, the
wavefunction and spectrum of {\em KK} modes is different
for $t_R$ as compared to $u_R$ and $c_R$ (cf. down type quarks).
Both these effects result in $O(1)$ flavor dependence in $a_u$.
Thus, NP contributions to both $t_R \rightarrow c_L \gamma$
and $t_L \rightarrow c_R \gamma$ from KK gluon exchange are comparable to
those from Higgs diagrams (unlike for down-type quarks).


\section{Higgs contributions to dipole operators}\label{ADipole}

The flavor structure of Higgs contributions to the dipole operators 
[discussed in subsection (\ref{Dipole})] are of two types.
One is mediated through both down and up type Yukwa couplings
while the other
proceeds only through down type Yukawa couplings.
Below we calculate the flavor structure of the induced operators
$$O_{\gamma}^{ij}=\bar d^i
_L \sigma_{\mu\nu}d_R^j F^{\mu\nu}\,,$$
where $F^{\mu\nu}$
is the field strength for the electromagnetic (or strong)
interaction
and $i,j=1..3$.
The Wilson coefficient of the corrresponding NP contribution,
$C_{\gamma}^{ij}$ (or $C_{8g}^{ij}$), are proportional to
\begin{eqnarray}
  C_{\gamma}^{ij}&\propto& 2 k^3 v F_Q  \left(
 \lambda_{u\, 5D }\lambda_{u\, 5D }^\dagger
+
\lambda_{d\, 5D }\lambda_{d\, 5D }^\dagger\right) \lambda_{d\, 5D } F_d\label{dog}
\,,\end{eqnarray}
where the above is in the interaction basis.\footnote{In the Higgs diagram
with up-type $5D$ Yukawa, there is additional flavor-violation from 
splitting in KK spectrum
for $u$ 
due to $c_{ u^3 } << c_{ u^{ 1,2 } }$. Whereas, 
$d$, $Q$ KK modes are almost degenerate (due to all $c$'s $\sim 1/2$) so 
that there is
no such effect in 
Higgs diagram with only down-type $5D$ Yukawa or
in the KK gluon diagram for down-type quarks as already mentioned.}
Before we make a detailed computation
we can check whether the above two contribution can in principle yield
 non-trivial flavor physics. This is the case if the two contrbutions
 are separately missaligned with the down type mass matrix
 $F_Q\lambda_{d\, 5D }F_d$.
 In order to check whether they are aligned 
we
calculated the
corresponding commutators
and found that, in general, the commutators are
non zero so that $C_{\gamma}^{ij}$ and $4D$ down-type Yukawa
cannot be
diagonalized simultanously.
 \subsection{Dipole operators - generic structure}\label{DipoleGeneric}

In order to compute the size of the new contribution to
$C_{\gamma}^{ij}$
we manipulate the above expression and rewrite it as a function of
SM fermion masses (in the mass basis):
\begin{eqnarray}
 \hspace*{-.4cm} C_{\gamma}^{ij}&\propto& 2 k^3 v
  \left[   F_Q\left(
 \lambda_{u\, 5D }\lambda_{u\, 5D }^\dagger
+
\lambda_{d\, 5D }\lambda_{d\, 5D }^\dagger\right) \lambda_{d\, 5D }
 F_d \right]_{ij}
\nonumber\\
&=& 2k^3 v \left[ F_Q\left(
 \lambda_{u\, 5D }\lambda_{u\, 5D }^\dagger
+
\lambda_{d\, 5D }\lambda_{d\, 5D }^\dagger\right) F_Q^{-1} 
F_Q\lambda_{d\, 5D }
 F_d \right]_{ij}\nonumber\\
&=& 
k^2 m_{d^j} \left[F_Q\left(
 \lambda_{u\, 5D }\lambda_{u\, 5D }^\dagger
+
\lambda_{d\, 5D }\lambda_{d\, 5D }^\dagger\right) F_Q^{-1} 
\right]_{ij}
\nonumber\\
&=& 
k^2 m_{d^j} \left[ \left(
 F_Q 
\lambda_{u\, 5D }F_u  F_u^{-1}\lambda_{u\, 5D }^\dagger
\right.\right.
\nonumber\\
&&\,\,\left.\left.  \ \ \ \ +
F_Q \lambda_{d\, 5D } F_d  F_d^{-1}
\lambda_{d\, 5D }^\dagger\right) F_Q^{-1} 
\right]_{ij}
\nonumber\\
&=& 
{m_{d^j}\over 4v^2} 
\left\{\left[
    V_{\rm CKM}^\dagger \,{\rm diag}\left( m_{u,c,t}\right)
  U_R^\dagger \,{\rm diag}\left( f^2_{u^i}\right)
  U_R \,{\rm diag}\left( m_{u,c,t}\right)
  U_L^\dagger
\right.\right.
\nonumber\\
&&\,\,\left.\left.  \ \ \ \ +
{m_{d^i}}  D_R^\dagger \,{\rm diag}\left( f^2_{d^i}\right) D_R  \,{\rm diag}\left( m_{d,s,b}\right)
D_L^\dagger \right] \,{\rm diag}\left( f^2_{Q^i}\right) D_L
\right\}_{ij}\hspace*{-.1cm},
\label{dipolegen}
\end{eqnarray}
where the first (second) term in each equality comes from the diagram
in fig. 6 (fig. 5) and involve both up and down (only down) KK quarks
and in the last line we give the explicit form of $F_Q F_Q^\dagger$
and $F_{u,d}^{\dagger}F_{u,d}$ in the mass basis.
Without loss of generality one can always go to a basis in which
$F_{Q,u,d}$ are real and diagonal.
In  this basis the down type contibutions, $C_\gamma^{d^{\rm KK}}$, are
proportional to:
\beq
\left(C_\gamma^{d^{\rm KK}}\right)_{ij}\propto
{m_{d^j} {m_{d^i}}\over 4v^2} 
\left[  R  \,{\rm diag\,}\left( m_{d,s,b}\right)
L
\right]_{ij}\,,
\label{dipolegend}
\end{eqnarray}
where 
$$\left({R,L}\right)_{ij}=\sum_n\left(D_{R,L}^*\right)_{ni}\left(D_{R,L}\right)_{nj}
f_{d^n,Q^n}^2\,.$$
One can convince himself that order of magnitude-wise
\beq\left(L,R\right)_{ij}\sim
\left(D_{R,L}^*\right)_{1i}\left(D_{R,L}\right)_{1j}
f_{d^1,Q^1}^2\,.  \label{Ldown}\eeq
Thus the elements of $R,L$ has the following hierarchy,
\beq
{(L,R)_{12}\over (L,R)_{11}}&\sim& {f_{Q^2,d^2}\over f_{Q^1,d^1}}\,,
\qquad 
{(L,R)_{22}\over (L,R)_{11}}\sim {f_{Q^2,d^2}^2\over f_{Q^1,d^1}^2}\,,
\qquad 
{(L,R)_{13}\over (L,R)_{11}}\sim {f_{Q^3,d^3} \over f_{Q^1,d^1}}\,,
\nonumber\\
{(L,R)_{23}\over (L,R)_{11}}&\sim &{f_{Q^2,d^2}f_{Q^3,d^3}\over
  f_{Q^1,d^1}^2}\,, \qquad
{(L,R)_{33}\over (L,R)_{11}}\sim {f_{Q^3,d^3}^2\over f_{Q^1,d^1}^2}\,.
\label{hierLR}
\eeq

The up type contibutions, $C_\gamma^{u^{\rm KK}}$, are
proportional to:
\beq
\left(C_\gamma^{u^{\rm KK}}\right)_{ij}\propto
{m_{d^j}\over 4v^2} 
\left[V_{\rm CKM}^\dagger  \,{\rm diag\,}\left( m_{u,c,t}\right)
  \bar R  \,{\rm diag\,}\left( m_{u,c,t}\right)
\bar L V_{\rm CKM}
\right]_{ij}\,,
\label{dipolegenu}
\end{eqnarray}
where 
$$\left({\bar R,\bar L}\right)_{ij}=\sum_n\left(U_{R,L}^*\right)_{ni}\left(U_{R,L}\right)_{nj}
f_{u^n,Q^n}^2\,,$$ and magnitude-wise we find
\beq\left(\bar L,\bar R\right)_{ij}\sim
\left(U_{R,L}^*\right)_{1i}\left(U_{R,L}\right)_{1j}
f_{u^1,Q^1}^2\,,  \label{Lup}\eeq
and the pattern of hierarchy between the elements of $\bar L,\bar R$
is similar to what shown in (\ref{hierLR}).

\subsection{Dipole operators and $b\to d,s\gamma$}\label{apphiggsbsg}
Using eqs. (\ref{dipolegend},\ref{dipolegenu}) we
can estimate, in term of spurions, what is the NP contribution,
$C_\gamma^{\lambda_d},C_\gamma^{\lambda_u}$,
to the opposite chiralty operator $C'_{7\gamma}\,.$

We first consider the contribution from the diagram in fig. 5 which
contains only down type KK quark.
The relevant NP contribution to $b\to s\gamma$ is proportional
$\left(C_\gamma^{d^{\rm KK}}\right)_{32}$ (\ref{dipolegend}),
    \beq
\hspace*{-.7cm}    C^{\lambda_d}_{7\gamma}&\propto&\left(C_\gamma^{d^{\rm
  KK}}\right)_{32}=
{m_{s} {m_{b}}\over 4v^2} 
\left[  R  \,{\rm diag\,}\left( m_{d,s,b}\right)
L
\right]_{32}\nonumber\\
&\sim& {m_{s} {m_{b}}\over 4v^2} \left(m_d R_{31} L_{12}+m_s R_{32}
  L_{22}+m_b R_{33} L_{32}\right)\nonumber\\
&\sim& {m_{s} {m_{b}}\over 4v^2}\, f_{d^3}
  f_{Q^2}\left( a_1 m_d {f_{Q^1}f_{d^1} }+a_2 m_s f_{d^2}
  f_{Q^2}+a_3 m_b {f_{d^3}
    f_{Q^3}}\right)\nonumber\\
&\sim& a_d { m_{s} {m_{b}}\over 4v^2}\,  f_{Q^2} f_{d^2} \left(D_R\right)_{23}
  2k v\lambda_{5D}\sim   \left(  \lambda_{ 5D } k \right)^2 { {m_{b}}} 
\left(D_R\right)_{23}
\label{bsgdown}
  \,,
  \eeq
  where $a_{i,d}$ is an  order one complex number.
  Similarly the contribution from the diagram in fig. 6 which
contains both down and up type KK quarks is proportional
$\left(C_\gamma^{u^{\rm KK}}\right)_{32}$ [see eq. (\ref{dipolegenu})],
    \beq
\hspace*{-.7cm}    C^{\lambda_u}_{7\gamma}&=&\left(C_\gamma^{u^{\rm
  KK}}\right)_{32}\propto
{m_{s} \over 4v^2} 
\left[V_{\rm CKM}^\dagger \,{\rm diag\,}
\left( m_{u,c,t}\right) \bar R  \,{\rm diag\,}\left( m_{u,c,t}\right)
\bar L V_{\rm CKM} \right]_{32}\nonumber\\
&\sim& a_u\,{m_{s} m_t m_u\over 4v^2}\, 
\, f_{u^3} f_{Q^2} f_{Q^1}f_{u^1}  
\sim a_u\,{m_{s} m_t \over 2v}\, 
k\lambda_{5D} f_{u^3} f_{Q^2}   
\nonumber\\
&\sim& a_u\,{m_{s}}\left(k\lambda_{5D}\right)^2 \,
{f_{Q^2}\over f_{Q^1}}\sim \left(  \lambda_{ 5D } k \right)^2 { {m_{b}}} 
\left(D_R\right)_{23}
\label{bsgup}
  \,,
  \eeq
where $a_{u}$ is an  order one complex number and 
in this case there
are nine terms of similar order so for simplicitly we represented them 
by a single term with a coefficient $a_u$.

Similar derivation yields the NP
contributions to the dipole moment operator $O^{\prime}_{7\gamma\; d}$
which mediates the $b\to d\gamma$ process:
\begin{equation}
C'_{7\gamma\; d}\propto
\left(  \lambda_{ 5D } k \right)^2 m_b \left( D_R \right)_{13}\,.\label{bdgam}
\end{equation}


\subsection{Flavor diagonal dipole operators}\label{Ddn}

In the above we showed explicitly that the above framework yields 
sizable contribution to the dipole operators 
$O'_{7\gamma},O^{\prime d}_{7\gamma}\,.$
In this part we ask whether similar contributions may yield
also contribution to flavor diagonal CPV observable of the forms
of neutron and electron EDMs.

In order to answer the above question we need to calculate
the relevant Wilson coefficient, $C_{d_n}$, that is generated
by the diagram in figures 5 and 6 (with $d_{L,R}$ external quarks).
We aim to demonstrate that a physical imaginary part of $C_{d_n}$
is generated from the new diagrams.
We shall see that the contribution we get is due to the presence of
the ``Majorana-like'' phases. In particular we find below
that it is enough to have two generations in order
to obtain the non-vanishing contributions.
For simplicity, to demonstrate our point, it is enough to consider only the contribution
with internal down quarks (figure 5). The diagram in figure 6 is
expected
to induce similar but independent contributions of roughly the same magnitude.
We start by using the result for $C^{ij}_{\gamma}$ (\ref{dipolegen})
with $i,j=1\,,$:
\begin{eqnarray}
C_{d_n}&\propto&
 {m_{d}^2\over 4v^2} 
\left[ \left(F_d^{\dagger}F_d\right)^{-1} 
\left( m_{d,s,b}\right)\left(F_Q F_Q^\dagger\right)^{-1} 
\right]_{11}\,,
\label{dn}
\end{eqnarray}
where the above result is in the mass basis.
It is clear that the above Wilson coefficient receive
non-zero contributions.
In order to have CPV, however, we should
check whether the above contributions
contains non removable CP phases, i.e, physical CPV phases.
Since the external quarks are in the mass basis the
only phase redefinition freedom allowed is a vector like rotation,
$d^i_{L,R}\to d^i_{L,R} e^{i\chi^i}\,.$
One can easily convince himself that this transformation leaves invariant
the 11 element of the object in the square paranethesis of (\ref{dn}).
Consequently $C_{d_n}$ cannot be brought to be real by such a simple
field redefinition.

In order to explicitly calculate the imaginary part of $C_{d_n}$
we look more closely at the expression in (\ref{dn}).
\begin{eqnarray}
  \hspace{-.2cm}C_{d^n}^1\propto
 \left[D_R^\dagger \,{\rm diag}\left( f^2_{d^1,d^2,d^3}\right)  D_R \,
   {\rm diag}\left(m_{d,s,b}\right)
   D_L^\dagger \,{\rm diag}\left( f^2_{Q^1,Q^2,Q^3}\right) D_L
\right]_{11}\,,
  \label{dn1gen1}
\end{eqnarray}
where this is in the ``special basis'' in which $F_{Q,d}$ are real 
and diagonal.

Let us make a short detour from the main route of the above discussion 
to see how this fits with our discussion in subsection (\ref{SpIm}). In
that part we showed that if
we have flavor violation only in the down type sector then $D_{L,R}$
contains 4 CPV phases.
In order to see how these are distributed
consider, {\it e.g.}, the following general parameterization of 
a $3\times 3$ complex down quark mass matrix, $M_d\,,$
\begin{eqnarray}
  M_d =D_L {\rm diag}\left(m_{d,s,b}\right)
  D_R^\dagger\,,
\label{Md}\end{eqnarray}
where $ D_{L,R}$ are $3\times3$ unitary matrices:
\begin{eqnarray}
  D_{R}^\dagger&=& P_D\,
R_{12}^{R} \,R_{13}^{R}\,{\rm diag}\left(1,1,e^{i\delta^{R}}\right) R_{23}^{R} 
\,{\rm diag}\left(1,e^{i\theta^{R}_1},e^{i\theta^{R}_2}\right)\,,
\nonumber\\
D_L^\dagger&=& 
R_{12}^{L}\, R_{13}^{L} \,{\rm diag}\left(1,1,e^{i\delta^{L}}\right)R_{23}^{L} 
\,{\rm diag}\left(1,e^{i\theta^{L}_1},e^{i\theta^{L}_2}\right)\,,
\label{rot1}
\end{eqnarray}
where $P_D={\rm diag}(e^{i\phi_1},e^{i\phi_2},e^{i\phi3})$
and $R_{ij}^{L,R}$ stands for an SO(3) matrix which describes a
rotation in the $ij$ plane and $\delta_{R,L}$ are CKM like phases
while the other are ``Majorana''
like.~\footnote{Note that the above mass matrix has
  9 idependent phases as required.}
As discussed in subsection~\ref{SpIm} we have, still in the
``special'', interaction, basis, a freedom to rotate five
phases using field redefinitions for the down quarks in the
interaction basis
$d_{L,R}^i\to d_{L,R}^i e^{i\chi_{L,R}^i}$ (in order to remain in
the ``special basis", in which $F_{Q,u,d}$ are real and diagonal,
all the KK excitation should also be similarly rotated). 
This will modify the form of the mass matrix which will contain
only 4 phases 
$M_d\to \tilde M_d={\tilde D_L}\, {\rm diag}\left(m_{d,s,b}\right)
  \tilde D_R^\dagger\,,$ where
\begin{eqnarray}
  \tilde D_{R}^\dagger&=& 
{\rm diag}\left(1,e^{i\tilde \phi_1},e^{i\tilde \phi_2}\right)
R_{12}^{R}\,{\rm diag}\left(1,1,e^{i\delta^{R}}\right) R_{13}^{R} R_{23}^{R} 
\,,
\nonumber\\
\tilde D_L^\dagger&=& 
R_{12}^{L}\,{\rm diag}\left(1,1,e^{i\delta^{L}}\right) R_{13}^{L} R_{23}^{L} 
\,,
\label{rot2}
\end{eqnarray}
where in this definition only $\tilde D_R$ contains, unremovable, Majorana type phases 
$\tilde\phi_{1,2}$. These can however be shifted
to $\tilde D_L$ using vector like field redefinitions in the mass basis.
Thus, with mixing, the product in Eq. (\ref{dn1gen1}) might have non 
zero imaginary part which appears when several flavors are involved in
a physical process 
(since the contribution of only $\left( D_{R, L} \right)_{11}$ in Eq. 
(\ref{dn1gen1}) is real).

Let us write the above expression more explicitly:
\begin{eqnarray}
  C_{d^n}^1
  \propto &&\left(R_{11},\,
{m_s\over m_d}R_{12},\,
{m_b\over m_d}R_{13}  \right)
\nonumber\\
&&\cdot
 \left(L_{11},\,
L_{12},\,
L_{13}  \right)^T
\,,\label{dn1genNM}
\end{eqnarray}
where 
$$\left({R,L}\right)_{ij}=\sum_n\left(D_{R,L}^*\right)_{ni}\left(D_{R,L}\right)_{nj}
f^2_{d^n,Q^n}$$
and the dot stands for a scalar product between the two vectors.
Note the in the above expression the first element of each vector is real thus
cannot contribute to the EDM.

To further simplify the analysis of (\ref{dn1genNM})
we move to a two generations framework. In that case we have a single,
unremovable, CPV ``Majorana''
like phase which again can be shifted from $D_L$ to $D_R\,$
\begin{eqnarray}
  \tilde D_{R}^\dagger&=& 
{\rm diag}\left(1,e^{i\tilde \phi}\right)
R_{12}^{R} 
\,,\qquad
\tilde D_L^\dagger=
R_{12}^{L} 
\,,
\label{rot3}
\end{eqnarray}
Then we find
\begin{eqnarray}
  C_{d^n}^1
  \propto R_{11}L_{11}+{m_s\over m_d} \left[e^{-i\tilde\phi}f_1^2
    \left|\left(D_R\right)_{11} \left(D_R\right)_{12}
          \right|+f_2^2
    \left(D_R\right)_{21} \left(D_R\right)_{22}
    \right] L_{12}
\,,\label{dn1genNMFin}
\end{eqnarray}
where note that the above expression is invariant with under 
vector-like field redefinitions in the mass
basis.
Thus the above does contribute to the EDM. Furthermore using
eq. (\ref{hierLR})
one can convince himself that the contribution is unsuppressed
since the suppression due to mixing is compensated by the $m_s/m_d$ enhancement so that
altogether we find
\begin{eqnarray}
  {\rm Im}\left(C_{d^n}^1\right)\propto m_d
\,.\label{dn1genNMFin1}
\end{eqnarray}
It is clear that also in the three generation case a similar result
is obtained. This is since the contribution is due to the non-removable Majorana
phases and we showed in the above that the resultant structure
is invariant with respect to field redefinitions.


\end{appendix}


\end{fmffile}
\end{document}